\documentclass[final, 5p, times, authoryear, twocolumn]{elsarticle}
\usepackage{todonotes}
\usepackage[most]{tcolorbox}

\usepackage{hyperref}
\usepackage{blindtext}

\usepackage{graphicx}
\usepackage{float}
\usepackage{diagbox}
\usepackage{epstopdf}
\usepackage{caption}
\usepackage{subfig}
\usepackage{xcolor}
\usepackage{pgfplots}
\usepackage{tikz}
\pgfplotsset{compat=1.10}
\usetikzlibrary{decorations.pathreplacing}

\usepackage{listings}
\usepackage{xcolor}

\definecolor{codegreen}{rgb}{0,0.6,0}
\definecolor{codegray}{rgb}{0.5,0.5,0.5}
\definecolor{codepurple}{rgb}{0.58,0,0.82}
\definecolor{backcolour}{rgb}{0.95,0.95,0.92}

\lstdefinestyle{mystyle}{
 backgroundcolor=\color{backcolour},  commentstyle=\color{codegreen},
 keywordstyle=\color{magenta},
 numberstyle=\tiny\color{codegray},
 stringstyle=\color{codepurple},
 basicstyle=\ttfamily\footnotesize,
 breakatwhitespace=false,     
 breaklines=true,         
 captionpos=b,          
 keepspaces=true,         
 numbers=left,          
 numbersep=5pt,         
 showspaces=false,        
 showstringspaces=false,
 showtabs=false,         
 tabsize=2
}

\date{ }

\usepackage{xcolor,soul}
\sethlcolor{lightgray}
\usepackage{tikz}

\usepackage[flushleft]{threeparttable}
\usepackage{tikz}
\usetikzlibrary{decorations.pathreplacing,calc}

\newcommand{\tikzmark}[2][-3pt]{\tikz[remember picture, overlay, baseline=-0.5ex]\node[#1](#2){};}

\tikzset{brace/.style={decorate, decoration={brace}},
 brace mirrored/.style={decorate, decoration={brace,mirror}},
}

\newcounter{brace}
\newcommand{\drawbrace}[3][brace]{
 \refstepcounter{brace}
 \tikz[remember picture, overlay]\draw[#1] (#2.center)--(#3.center)node[pos=0.5, name=brace-\thebrace]{};
}

\newcounter{arrow}
\newcommand{\annote}[3][]{
 \tikz[remember picture, overlay]\node[#1] at (#2) {#3};
}

\begin{document}

\begin{frontmatter}

 \title{A Fly in the Ointment: An Empirical Study on the Characteristics of Ethereum Smart Contracts Code Weaknesses and Vulnerabilities}

\author[1]{Majd Soud\corref{cor1}}
\ead{majd18@ru.is}
\author[1]{Grischa Liebel}
\ead{grischal@ru.is}
\author[1,2]{Mohammad Hamdaqa}
\ead{mhamdaqa@polymtl.ca}

\cortext[cor1]{Corresponding author}
\address[1]{School of Computer Science, Reykjavik University, Reykjavik, Iceland}
\address[2]{Department of Computer and Software Engineering, Polytechnique Montreal, Montreal, Canada}

\begin{abstract}
\textbf{Context: }Smart contracts are computer programs that are automatically executed on the blockchain. Among other issues, vulnerabilities in their implementation have led to severe loss and theft of cryptocurrency. In contrast to traditional software, smart contracts become immutable when deployed to the Ethereum blockchain. Therefore, it is essential to understand the nature of vulnerabilities in Ethereum smart contracts to prevent these vulnerabilities in the future. Existing classifications exist, but are limited in several ways, e.g., focusing on single data sources, mixing dimensions, or providing categories that are not orthogonal.\\
\textbf{Objective:} This study aims to characterize vulnerabilities and code weaknesses in Ethereum smart contracts written in Solidity, and to unify existing classifications schemes on Ethereum smart contract vulnerabilities by mapping them to our classification. \\
\textbf{Method:} We extracted 2143 vulnerabilities from public coding platforms (i.e., GitHub and Stack Overflow) and popular vulnerability databases (i.e., National Vulnerability Database and Smart Contract Weakness Registry) and categorized them using a card sorting approach. We targeted the Ethereum blockchain in this paper, as it is the first and most popular blockchain to support the deployment of smart contracts, and Solidity as the most widely used language to implement smart contracts.
We devised a classification scheme of smart contract vulnerabilities according to their error source and impact. Afterwards, we mapped existing classification schemes to our classification.\\
\textbf{Results:} The resulting classification consists of 11 categories describing the error source of a vulnerability and 13 categories describing potential impacts. Our findings show that the language specific coding and the structural data flow categories are the dominant categories, but that the frequency of occurrence differs substantially between the data sources.\\
\textbf{Conclusions:} Our findings enable researchers to better understand smart contract vulnerabilities by defining various dimensions of the problem and supporting our classification with mappings with literature-based classifications and frequency distributions of the defined categories. Also, they allow researchers to target their research and tool development to better support the implementation and quality assurance of smart contracts.
\end{abstract}


\end{frontmatter}

\section{Introduction}

To process autonomous tasks, some blockchains execute digital programs called \emph{smart contracts} that become immutable once deployed \citep{zheng2020overview}. 
Smart contracts are autonomous programs that can (1) customize contracting rules and functions between contractors and (2) facilitate transferring irreversible and traceable digital cryptocurrency transactions \citep{hewa2020survey}. Smart contracts were born and continued to grow with noteworthy achievements in various life fields, including industry, finance, and economic.

Due to the immaturity of blockchain technology, vulnerabilities in smart contracts can have severe consequences and result in substantial financial losses. 
For instance, the infamous DAO attack in 2016 \citep{Dao1} resulted in stealing around 50 million dollars because of exploiting a re-entrancy vulnerability in the Distributed Autonomous Organizations (DAO) contract\footnote{https://www.coindesk.com/understanding-dao-hack-journalists}.

Therefore, understanding vulnerabilities in smart contracts is critical to perceive the threats they represent, e.g., to develop predictive models or software engineering tools that can predict or detect threats with a high precision \citep{seacord2005structured}.
Furthermore, classifying smart contract vulnerabilities enables researchers and practitioners to better understand their frequency and trends over time

Existing studies attempt to categorize vulnerabilities in Ethereum smart contracts~\cite{atzei2017survey,dingman2019defects,chen2020defining,zhang2020framework,rameder2021systematic}.
While they provide valuable insights into existing security issues in smart contracts, they fall short in several ways:
\begin{enumerate}
    \item \textbf{There is no unified view on smart contract vulnerabilities.} For instance, \citep{atzei2017survey} classify security vulnerabilities according to their network level, \citep{chen2020defining} classify defects in smart contracts according to their impact on quality attributes, and \citep{zhang2020framework} categorize smart contract defects according to sources of error. While these dimensions are all relevant, they are orthogonal and cannot easily be compared.
    \item \textbf{Several studies mix different classification dimensions.} For instance, \citep{zhang2020framework} focus primarily on error sources (e.g., data and interface errors), but also include categories concerned with effects/impact on quality attributes (e.g., security and performance). This leads to classifications where categories are not orthogonal and, since these dimensions are not discussed, often confusing.
    \item \textbf{Vulnerabilities are classified into broad categories.} Several studies classify vulnerabilities into broad categories, which results in two main shortcomings. First, a vulnerability can be assigned into more than one category. Second, the differences among the categories are too general to be useful to reason about the vulnerabilities. 
    \item \textbf{Data sources differ widely.} Several existing studies rely only on vulnerabilities published in academic literature or white literature, e.g., \citep{atzei2016survey} and \citep{alharby2017blockchain}, while \citep{chen2020defining} used posts on StackExchange, and \citep{zhang2020framework} used a mix of academic literature and GitHub project data. This makes the comparison challenging or even impossible.
    \item \textbf{Important data sources are omitted in existing classifications.} To our knowledge, no existing study uses established vulnerability and defect registries such as the Smart Contract Weakness Registry (SWC) \footnote{https://swcregistry.io/} and vulnerabilities in the Common Vulnerability and Exposure (CVE)\footnote{https://cve.mitre.org/}.
\end{enumerate}

To address these gaps, the contribution of this study is twofold:
\begin{enumerate}
    \item To unify existing classifications on smart contract vulnerabilities by providing an overview of the different classification dimensions, and by mapping existing classifications to a single classification scheme using error source and impact as dimensions of the vulnerabilities. 
    \item To complement existing studies by classifying smart contract vulnerabilities extracted from a variety of important data sources according to the different dimensions presented in existing work.
    \end{enumerate}

We extracted and analyzed data related to Ethereum smart contracts written in Solidity from four data sources, i.e., Stack Overflow, GitHub, CVE and SWC. Using a card sorting approach, we devised a classification scheme that uses the error source and impact of a vulnerability as dimensions. Furthermore, we mapped existing classfications to this scheme and analyzed the frequency distribution of the defined categories per data source.
The resulting classification scheme consists of 11 categories describing the error source, and 13 categories describing potential impacts.

Our findings show that language specific coding and structural data flow categories are the dominant categories of vulnerabilities in Ethereum smart contracts. However, frequency distribution of the error source categories differ widely across data sources. With respect to the existing classifcations, we find that the majority of sources use broad categories that are applicable to many vulnerabilities, such as ``security'' or ``availability''.

The remainder of this paper is organized as follows. Section~\ref{sec:background}
presents the background on Ethereum smart contracts, Solidity and the used data sources. Section~\ref{sec:related_work} discusses how existing work relates to our study and what gaps exist. Section~\ref{fig:method} describes the methodology that we followed. In Section~\ref{sec:results}, we report our findings in terms of the obtained classification and mapping to existing work. We then discuss our findings in Section~\ref{sec:discussion} and potential threats to validity in Section~\ref{sec:threats}. The paper is concluded in Section~\ref{sec:conclusion}.

\section{Background}
\label{sec:background}
 In this section, we discuss the background of our work.
 Specifically, we discuss definitions used in this paper, the Ethereum environment, and existing vulnerability and weakness databases.

\subsection{Definitions}
Before we can discuss a classification scheme for smart contracts' vulnerabilities, it is fundamental that we have a sufficiently specific definition of what we are classifying. There have been efforts to formally
define concepts such as vulnerability in SE. However, we will use the formal definition of a vulnerability and weakness that were defined in the Ethereum Improvement Proposals (EIPs)\footnote{https://eips.ethereum.org/} as follows:

\begin{itemize}
    \item \textbf{Vulnerability:} ``A weakness or multiple weaknesses which directly or indirectly lead to an undesirable state in a smart contract system'' (\cite{EIP1470} 1470). A vulnerable contract does not necessarily imply exploited \citep{perez2019smart}. Moreover, as the contract is a digital agreement between two or more parties, exploiting the contract is not always done by external malicious actors. A venerable contract can also be exploited by one of the contracting parties such as the contract owner who can use vulnerabilities to gain more profits such as in CVE-2018-13783\footnote{https://nvd.nist.gov/vuln/detail/CVE-2018-13783}. Any of the contractors can also exploit it, the miners or even the developers who implemented the contract (e.g., CVE-2018-17968 \footnote{https://nvd.nist.gov/vuln/detail/CVE-2018-17968}).

    \item \textbf{Weakness:}
     ``a software error or mistake in contract code that in the right conditions can by itself or coupled with other weaknesses lead to a vulnerability''. (\cite{EIP1470} 1470). 
    
\end{itemize}

What distinguishes smart contract code weaknesses from other software applications is that any smart contract code instruction costs a specific amount of gas (see Section-\ref{systemgas}). This means even if the weakness was not exploited, it would result in losing Ether \footnote{Ethereum corresponding cryptocurrency} when it is triggered by the contract itself and executed, which makes the contract vulnerable even if it is not exploited. Thus, it is of high importance to study smart contracts' weaknesses along with smart contracts' vulnerabilities. 

\subsection{\textbf{Ethereum Virtual Machine (EVM)}} Ethereum\footnote{https://ethereum.org/en/} is a globally open decentralized blockchain framework that supports smart contracts, referred to as Ethereum Virtual Machine (EVM). Because EVM hosts and executes smart contracts, it is often referred to as the programmable blockchain. EVM contracts reside on the blockchain in a Turing complete bytecode language; however, they are implemented by developers using high-level languages such as Solidity or Vyper and then compiled to bytecode to be uploaded to the EVM. Users on the EVM can create new contracts, invoke methods in a contract, and transfer Ether. All of the transactions on EVM are recorded publicly and their sequence determines the state of each contract and the balance of each user. In order to ensure the correct execution of smart contracts, EVM relies on a large network of mutually untrusted peers (miners) to process the transactions. EVM also uses the Proof-of-Work (PoW) consensus protocol to ensure that a trustworthy third party (e.g., banks) is not needed to validate transactions, fostering trust among users to build a dependable transaction ledger. EVM gained remarkable popularity among blockchain users as EVM is the first framework that supports smart contracts to manage digital assets and build decentralized applications (DApps) \citep{khan2020ethereum}.

\subsection{\textbf{Ethereum Smart Contracts}}
A smart contract is a general-purpose digital program that can be deployed and executed on the blockchain. Ethereum smart contract is identified by a unique 160-bit hexadecimal string which is the contract address. It is written in a high-level language, either Solidity or Vyper. In this paper, we focus on Ethereum smart contracts written in Solidity because it is the most popular language in the EVM community and most of the deployed contracts on EVM are written using Solidity \citep{bhat2017probabilistic,badawi2020cryptocurrencies}.
A smart contract can call other accounts, as well as other contracts on the EVM. For example, it can call a function in another contract and send Ether to a user account. In EVM, internal transactions (i.e., calls from within a smart contract) do not create new transactions and are therefore not directly recorded on-chain.

\subsection{\textbf{Ethereum Gas System}}
\label{systemgas}
In order to execute a smart contract, a user has to send a transaction (i.e., make a function call) to the target contract and pay a transaction fee that is measured in units of gas, referred to as the gas usage of a transaction. The transaction fee is derived from the contract’s computational cost, i.e., the type and the number of executed instructions during runtime. Each executed instruction in the contract consumes an agreed-upon amount of gas. Instructions that need more computational resources cost more gas than instructions that need less computational resources. This helps secure the system against denial-of-service attacks and prevents flooding the network. Hence, the gas system in EVM has two main benefits: (1) it motivates developers to implement efficient applications and smart contracts and (2) it compensates miners who are validating transactions and executing the needed operations for their contributed computing resources. To pay for gas, the transaction fee equals the \begin{math}gasprice * gascost\end{math}. The minimum unit of gas price is Wei (1 Ether = \begin{math} 10^{18} \end{math} Wei). Therefore, Ether can be thought of as the fuel for operating Ethereum. 

\subsection{\textbf{Solidity}}
Solidity\footnote{https://docs.soliditylang.org/} is a domain-specific language (DSL) that is used to implement smart contracts on the Ethereum blockchain. 
It is the most widely used open-source programming language in implementing blockchain and smart contracts. Also, although it was originally designed to be used on Ethereum, it can be used in other blockchain platforms such as Hyperledger and Monax\footnote{https://monax.io/}. Solidity is statically typed, which requires specifying the type of all variables in the contract. It does not support any ``undefined'' or ``null'' values, and any newly defined variable has a default value based on its type. Smart contracts written in Solidity are organized in terms of subcontracts, interfaces, and libraries. They may contain state variables, functions, function modifiers, events, struct types, and enum types. Also, Solidity contracts can inherit from other contracts, and can call other contracts.

In Solidity there are two kinds of function calls. \textit{Internal function calls}, which do not create an actual EVM call, and \textit{External function calls}, which do.
Due to this distinction, Solidity supports four types of visibility for functions and state variables:
\begin{itemize}
 \item \textit{External} functions can be called from other contracts using transactions as they are part of the contract interface. They can not be called internally (e.g., \emph{externalFunction()} does not work, while \emph{this.externalFunction()} works). State variables can not be external.
 \item \textit{Public} functions can be called internally or using messages and they are part of the contract interface. For public state variables, an automatic getter function is generated by the Solidity compiler to avoid high gas cost when returning an entire array. 
 \item \textit{Internal} functions and state variables can only be accessed internally (i.e., from within the current contract or contracts deriving from it).
 \item \textit{Private} functions and state variables are only visible for the contract they are defined in and not in derived contracts.
\end{itemize}

\begin{figure}[ht]
\centering
\includegraphics[width=.5\textwidth, bb=0 0 678 1091]{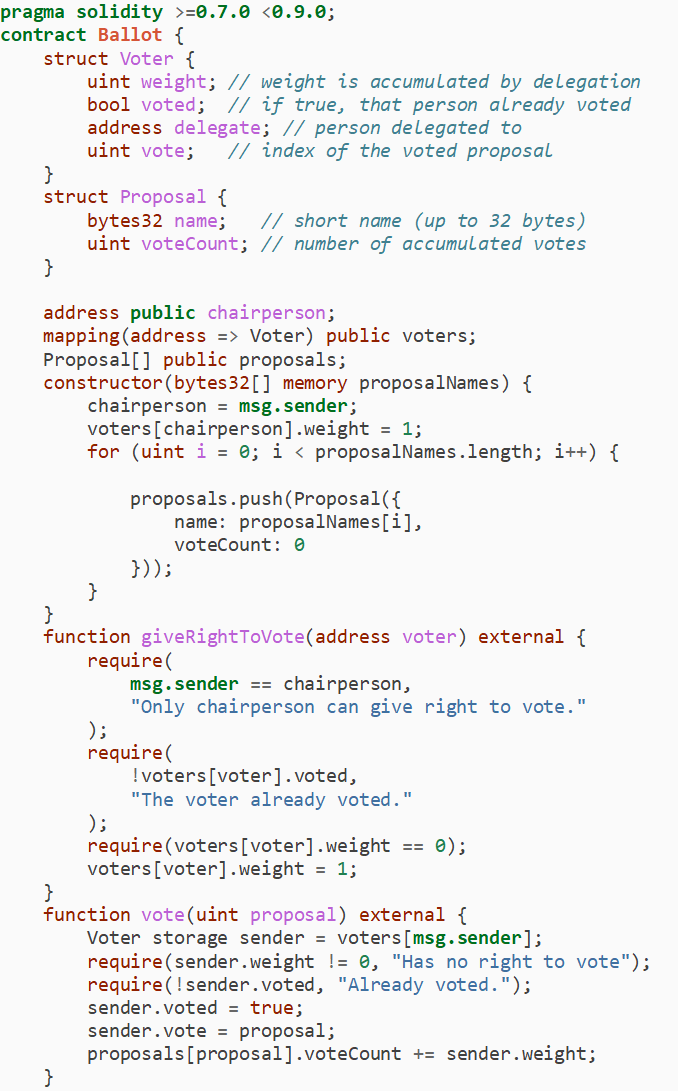}
\caption{Voting Contract Example Written in Solidity \citep{Soliditydocumentation}}
\label{fig:SolidityVotingContract}
\end{figure}

State variables can be declared using the keywords \textit{constant} or \textit{immutable}. Immutable variables can be assigned at construction time, while constant variables must be fixed at compile time. For the functions declaration, they can be declared as follows. 
\begin{itemize}
 \item \textit{Pure Functions} promise not to read or modify the state. They can use the \textit{revert()} and \textit{require()} functions to revert potential state changes when an error occurs.
 \item \textit{View Functions} promise not to modify the state such as writing to state variables.
 \item \textit{Receive Ether Functions} can exist at most once in a contract. They cannot have any arguments and cannot return any value, and must be declared with external visibility and payable state mutability as in \textit{receive() external payable\{ ...\}}. These functions are executed on plain Ether transfers (e.g., using \textit{.send()} or \textit{.transfer()}) and based on a call to the contract with empty \textit{calldata}.
 \item \textit{Fallback Functions} are similar to \textit{Receive Ether} functions, as any contract can have at most one such function. It must have external visibility, and is executed on a call to the contract if none of the other functions match the given function signature or if no data was supplied at all and there is no receive function. The \textit{fallback} function always receives data, but in order to also receive Ether it must be marked as \textit{payable}.
\end{itemize}
Finally, if any function promises to receive Ether, it has to be declared as \textit{payable}. An example of a Solidity contract is shown in Figure~\ref{fig:SolidityVotingContract}. It shows a voting contract as explained in Solidity's official documentation \citep{Soliditydocumentation}. 

\subsection{\textbf{Common Vulnerability and Exposure (CVE) and National Vulnerability Database (NVD)}}
CVE is a list of publicly disclosed vulnerabilities and exposures that is maintained by MITRE\footnote{https://cve.mitre.org/cve/}. It feeds into the NVD\footnote{http://nvd.nist.gov/}, so both are synchronized at all times. NVD is a comprehensive repository with information about all publicly known software vulnerabilities and includes all public sources of vulnerabilities (e.g., alerts from security focus\footnote{https://www.securityfocus.com/}). NVD also provides more information about the CVE lists' vulnerabilities, such as severity scores and patch availability. It also provides an easy mechanism to search for vulnerabilities using various variables. Both CVE and NVD are maintained by the US Federal Government. An example of a reported smart contract vulnerability in NVD is shown in Table~\ref{table:cveExample}. It also shows the impact metrics (Common Vulnerability Scoring System - CVSS), vulnerability types (Common Weakness Enumeration - CWE), applicability statements (Common Platform Enumeration - CPE), and other relevant meta-data. Each CVE in the list also has a unique identifier that shows the affected software product, sub-products, and various versions. 

\begin{table}[htbp]
\centering
\caption{An example of a vulnerability in NVD}
\label{table:cveExample}
\begin{tabular}{|p{3.4 cm} |p{4.4 cm}|} 
 \hline
 \hline
 CVE ID & CVE-2021-3006\\
NVD Published Date: & 01/03/2021 \\
Source: & CVE MITRE \\
Description & The breed function in the smart contract implementation for Farm in Seal Finance (Seal), an Ethereum token, lacks access control and thus allows price manipulation, as exploited in the wild in December 2020 and January 2021. \\
CVSS & 7.5 \\
Weakness Enumeration & CWE-863 \\
CWE Name & 	Incorrect Authorization \\
Hyperlink & Link\footnote{https://etherscan.io/address/0x33c2da7fd5b125e629b3950f3c38d7f721d7b30d} \\
Integrity & High\\
Impact score & 3.6 \\
\hline
\hline
\end{tabular}
\end{table}

\subsection{\textbf{Common Weakness Enumeration (CWE) }}
CWE \footnote{http://cwe.mitre.org} is a community-developed list of common software security weaknesses. It is considered as a comprehensive online dictionary of weaknesses that have been found in computer software. It also serves as a baseline for weakness identification, mitigation, and prevention efforts. Its primary purpose is to promote the effective use of tools to identify, find, and repair vulnerabilities and exposures in computer software before the programs are distributed to the public.

\subsection{\textbf{Smart Contract Weakness Classification Registry (SWC)}}
SWC is an implementation of the weakness classification scheme that is proposed in \textit{Ethereum Improvement Proposals}\footnote{https://eips.ethereum.org/}. It is also aligned with the terms and structures described in the CWE. Each SWC has an identifier (ID), weakness title, CWE parent, and related code samples list. 
\section{Related Work}
\label{sec:related_work}
Multiple studies that classify smart contract vulnerabilities have been published since the first attack on Ethereum smart contracts in 2016 \citep{daian2016analysis}, e.g., \cite{atzei2016survey}, \cite{alharby2018blockchain}, \cite{dingman2019defects}, and \cite{zhang2020framework}. This section summarizes these studies in relation to our work. 

\subsection{Literature-Based Vulnerability Classification}
Several studies on Ethereum smart contract vulnerabilities classify vulnerabilities reported in blogs, academic literature, and white papers, i.e., \cite{atzei2016survey,alharby2017blockchain,huang2019smart,sanchez2018raziel,chen2020survey,praitheeshan2019security,dingman2019defects}.

\cite{atzei2016survey} propose a classification scheme comprised of three categories to classify security vulnerabilities in Ethereum smart contracts. In addition to blogs and academic literature, the authors also employ their own practical experience as a resource for their classification. Vulnerabilities are classified into language-related issues, blockchain issues, and EVM bytecode issues. The classification is followed by a brief discussion on potential attacks that result in stealing money or causing other damage.

\cite{alharby2017blockchain} also use academic literature as the main source of data, classifying them into four main classes: codifying, security, privacy, and performance issues. Additionally, they discuss proposed solutions based on suggestions provided by smart contract analysis tools.
The proposed classification suffers from a significant overlap between categories. For example, codifying issues can lead to security and privacy issues, as in the case of the popular re-entrancy vulnerability (classified as a security issue in the paper).

\cite{huang2019smart} report a literature review of smart contract security from a software lifecycle perspective. The authors analyze blockchain features that may result in security issues in smart contracts and summarize popular smart contracts’ vulnerabilities based on four development phases, i.e., design, implementation, testing before deployment, and monitoring and analysis. Finally, they classified 10 vulnerabilities into three broad categories (i.e., Solidity, blockchain and misunderstanding of common practices). Unfortunately, there is no explanation of how or on what basis these categories were designed. 

\cite{dingman2019defects} study well-known vulnerabilities reported in white and gray literature and classify them according to National Institute of Standards and Technologies Bugs framework (NIST-BF)\footnote{https://samate.nist.gov/BF/} into security, functional, developmental, and operational vulnerabilities. The results show that the majority of vulnerabilities fall outside the scope of any category. However, the categories and the classification process are not defined or described in the paper.

Similar to \cite{dingman2019defects}, \cite{samreen2021survey} survey and map eight popular smart contracts’ vulnerabilities in the literature to the NIST-BF. Their results show that only three of the studied eight vulnerabilities could be matched with two NIST-BF classes. They also suggest a preventive technique per classified vulnerability. Finally, a map between existing analysis tools and the eight vulnerabilities are provided in the paper.  

\cite{praitheeshan2019security} classify 16 smart contracts’ vulnerabilities reported in literature based on their internal mechanisms. The authors use three categories, i.e., blockchain, software security issues, and Ethereum and Solidity vulnerabilities.

Finally, \cite{chen2020survey} survey Ethereum System Vulnerabilities including smart contract vulnerabilities reported in literature, classifying them according to two dimensions. First, they group vulnerabilities into four-layer groups according to the location, i.e., on the application, data, consensus or network layer. Secondly, they group vulnerabilities according to their cause into Ethereum design and implementation, smart contract programming, Solidity language and tool-chain, and human factors. This classification focuses on few locations that a vulnerability might occur at, while omitting others. For instance, the vulnerability could be located in the source code of the smart contract itself or in its dependencies. Furthermore, there is no clear indication of how these categories were defined or if a systematic way to classify them was followed.

\subsection{Repository-Based Vulnerability Classification}
In addition to classifications that are based on published vulnerabilities, two papers attempt classifications based on data extracted from public repositories such as StackExchange, i.e., \cite{chen2020defining,zhang2020framework}.

\cite{chen2020defining} collect smart contract vulnerabilities from discussions available on Ethereum StackExchange, classifying them based on five high-level aspects according to their consequences, i.e., security, availability, performance, maintainability, and reusability.
To evaluate if the selected vulnerabilities are harmful, the authors conduct an online survey to collect feedback from practitioners.
The proposed categories have the drawback that not all vulnerabilities can be clearly placed in a single category, i.e., one vulnerability could have various consequences. 
 
\cite{zhang2020framework} classify smart contracts vulnerabilities from both literature and open projects on GitHub. The authors classify extracted vulnerabilities into 9 categories based on an extension of IEEE Standard Classification for Software Anomalies\footnote{https://standards.ieee.org/standard/1044-2009.html}. Finally, they propose a four-category classification scheme for the impact of a vulnerability, i.e., unwanted function executed, performance, security, and serviceability. 
This classification is based on vulnerability GitHub reports, and some categories are too general to be useful in any detailed engineering analysis (e.g., the security category).

\subsection{Tool-Based Vulnerability Detection}
As a last category of related work, several publications study existing tools to detect vulnerabilities in smart contracts, and propose classifications based on the tools' capabilities, i.e., \cite{khan2020survey,rameder2021systematic}.

\cite{khan2020survey} provide an overview of current smart contracts vulnerabilities and testing tools in their work. In addition, they propose a vulnerability taxonomy for smart contracts' vulnerabilities. The proposed taxonomy consists of seven categories, i.e., inter-contractual vulnerabilities, contractual vulnerabilities, integer bugs, gas-related issues, transnational vulnerabilities, deprecated vulnerabilities, and randomization vulnerabilities. The authors then provide a mapping between the surveyed vulnerabilities and the available detection tools. Unfortunately, this categorization is not actually a classification scheme, in the sense that it fails to identify a category unique to each vulnerability and no structured method was provided to show how the classification was performed.

In a Master Thesis, \cite{rameder2021systematic} provides a comprehensive overview of state-of-the-art tools that analyze Ethereum smart contracts with an overview of known security vulnerabilities and the available classification schemes in the literature. The studied vulnerabilities are classified into 10 novel categories, i.e., malicious environment, environment dependency/blockchain, exception \& error handling disorders, denial of service, gas related issues, authentication, arithmetic bugs, bad coding quality, environment configuration, and deprecated vulnerabilities. 
However, the proposed categories cover different dimensions, e.g., vulnerability consequences as well as programming errors. Finally, some categories are not defined in the work. 

\subsection{Summary and Research Gap}
In summary, various attempts to classify smart contract vulnerabilities have been published, both on reported vulnerabilities and by mining software repositories.
A third line of research focuses on studying smart contract vulnerability detection tools, classifying what vulnerabilities they are able to detect.
These papers suffer from three flaws.
First, they rely exclusively on vulnerabilities reported in literature and might, therefore, provide a skewed image, e.g., \cite{atzei2016survey}.
Secondly, they propose classifications that mix different concerns or dimensions, such as consequences of exploiting a vulnerability and the source of error in \cite{rameder2021systematic}.
Third, they use categories with too broad distinctions that do not allow for detailed reasoning, such as the privacy and security categories in \cite{alharby2017blockchain} and \cite{zhang2020framework}.
Finally, they provide only a limited view on smart contract vulnerabilities due to focusing on a single dimension, such as the consequences in \cite{chen2020survey}.
Therefore, there is a need to unify existing taxonomies and classification schemes and provide a reference taxonomy that includes several dimensions such as root cause, impact, or scope \cite{vacca2021systematic}.
The aim of this paper is to arrive at such a classification by integrating existing work and complementing it with additional data from software repositories and well-known sources such as the CVE and SWC registries.

\section{Research Method}
\label{sec:method}
\begin{figure*}[htbp]
\centering
\includegraphics[width=0.94\textwidth, bb=0 0 791 518]{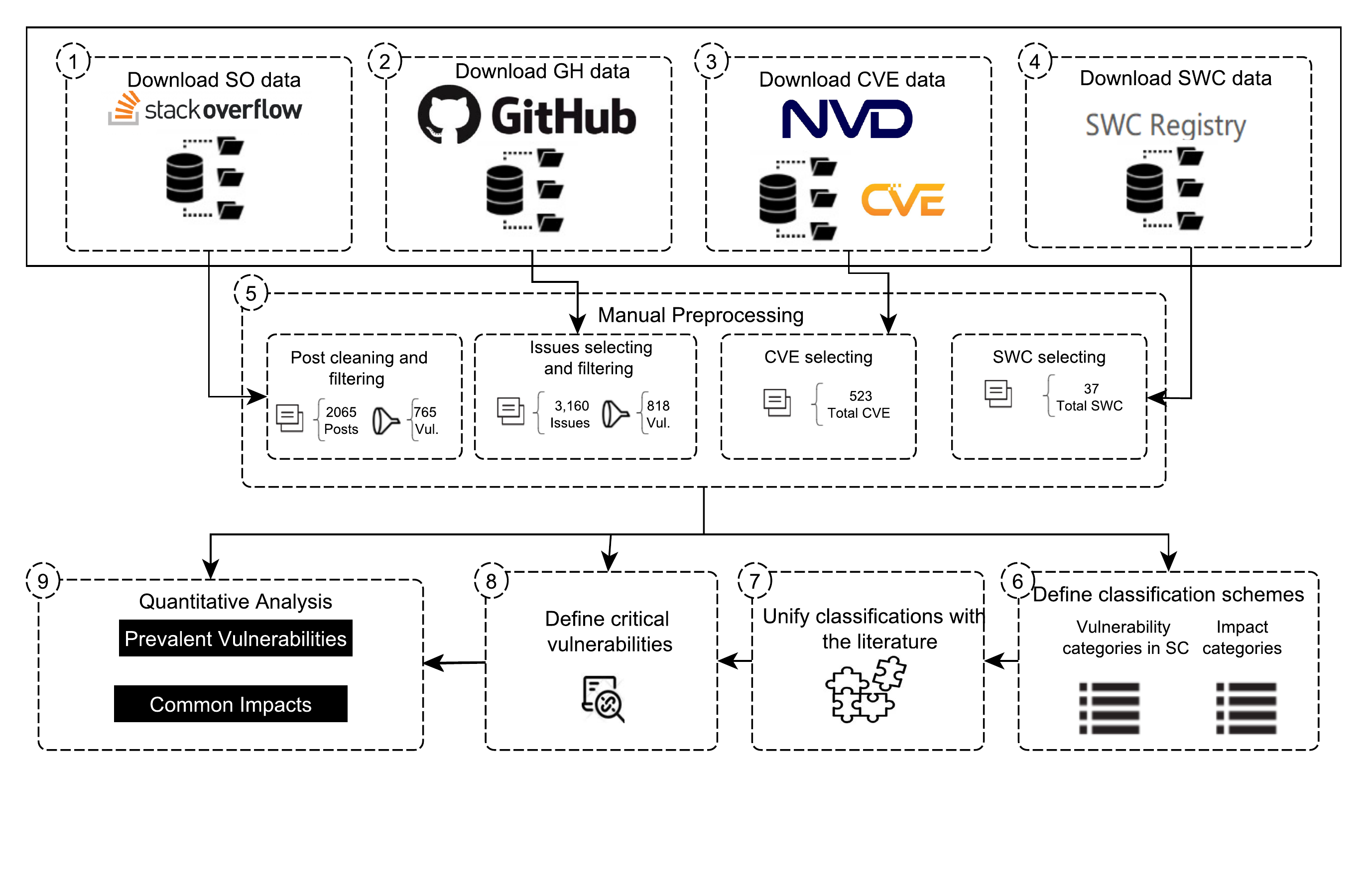}
\caption{Empirical Study 
 Method}
\label{fig:method}
\end{figure*}

This section presents the method we followed in this paper, as shown in Figure~\ref{fig:method}. It includes the study setup, data sources, data cleaning, and data analysis.

\subsection{Study setup}
In this study, we aim to answer the following research questions (RQs):
\begin{itemize}
\item \emph{RQ1. What categories of vulnerabilities appear in smart contracts?} 

Goal: To comprehensively categorize the vulnerabilities that appear in Ethereum smart contracts. To study different dimensions of the problem and to map literature-based classifications on Ethereum smart contract vulnerabilities and unify them in one thorough classification. A thorough classification scheme will make it possible to collect statistics on frequency, trends, and vulnerabilities, as well as evaluate countermeasures.

\item \emph{RQ2. Are the frequency distributions of smart contract vulnerability categories similar across all studied data sources?}

Goal: To investigate if all data sources have the same frequency distributions of vulnerabilities. If so, then we can rank vulnerability categories from the most common to the least common. If not, we can reason about the skew or bias when different sources are used. Further research can find solutions for the most common vulnerabilities. More effort could be put to address the prevalent category before deploying the contract to the blockchain.

\item \emph{RQ3. What impact do the different categories of smart contract vulnerabilities and weaknesses have?}

Goal: To investigate the impacts of smart contract vulnerability and code weaknesses. To define various dimensions of impacts classifications and to map literature-based impact classifications and propose a thorough impact classification of smart contracts vulnerabilities and code weaknesses. More effort can be put to vulnerabilities and code weaknesses with critical impacts.

\end{itemize}

\subsubsection{Data Sources}
To answer the proposed research questions, we analyzed and studied smart contract code vulnerabilities and weaknesses from four primary sources. Two of these sources are widely used by developers (i.e., Github and StackOverflow), and two are very well-known publicly accessible sources (i.e., SWC and NVD) for reporting vulnerabilities and weakness in Ethereum smart contracts and other software systems. Table~\ref{table:datasetSummary} shows the final number of data records during and after data pre-processing.

\begin{table}[htbp]
\centering
\caption{Summary of the sampled vulnerabilities in the selected data sources}
\label{table:datasetSummary}
\begin{tabular}{||p{2.6 cm} |p{1.4 cm} |p{0.9 cm} | p{0.6 cm} | p{0.8 cm}||} 
 \hline
 Data Source & Stack Overflow & GitHub & CVE & SWC\\ [0.5ex] 
 \hline\hline
 \# of collected data& 2065 & 3160 &523 & 37 \\
 \hline
 \# of data records after preprocessing steps& 1490 & 1160 &523 & 37 \\
\hline
 Final \# of Vulnerabilities& 765 & 818 &523 & 37 \\
\hline
\end{tabular}
\end{table}

\textbf{Data-source 1:} We opted for extracting data from Stack Overflow, as it has successfully been leveraged in existing work on smart contracts~\citep{ayman2019smart, aymansmart, chen2020defining}, and in general Software Engineering research \citep{bajaj14,ponzanelli14,calefato15,chen16,ahasanuzzaman16}.

We used Stack Overflow posts to study weaknesses and vulnerabilities in smart contracts.
To do so, we extracted Q\&A posts tagged with ``smart contract'', ``Solidity'', and ``Ethereum'', posted between January 2015 and April 2021.
We discarded posts with the ``Ethereum'' tag, but without the ``Solidity'' or ``smart contract'' tags.
To retrieve the related information for each post, we used Scrapy~\cite{Scrapy}, an open source Python library that facilitates Web crawling.
For each of the 2065 extracted posts, we extracted the post title, URL, related tags, post time and accepted answer time.

\begin{figure}[ht]
\centering
\includegraphics[width=.4\textwidth, bb=0 0 908 640]{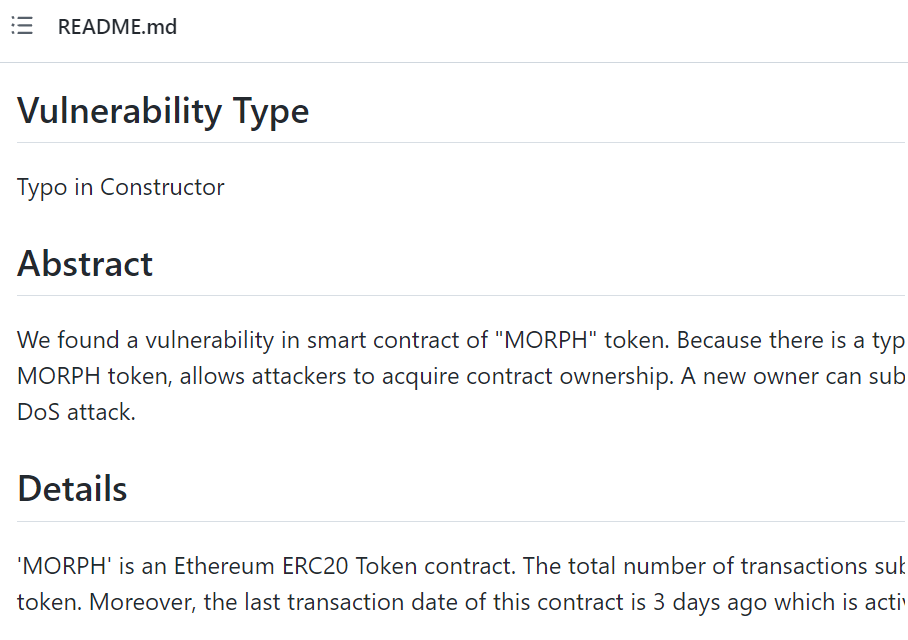}
\caption{Vulnerability Example from GitHub}
\label{fig:Githubvul}
\end{figure}

\textbf{Data-source 2:} We used GitHub as the second data source for our study, as it is the most popular social coding platform \citep{cosentino2017systematic}. Moreover, many studies on Ethereum smart contracts and smart contract analysis tools have been published reporting findings based on data published in GitHub open-source projects (e.g., \cite{durieux2020empirical}). We studied vulnerabilities and code weaknesses reported in open source projects that have Ethereum smart contracts written in Solidity. We used the keywords ``smart contract'', ``Solidity'', and ``Ethereum'' to search for these projects. Then we searched for the vulnerabilities and weaknesses based on the fixes in each project and based on the keywords ``Vulnerability'', ``bug'', ``defect'', ``issue'', ``problem'', and ``weakness''.

We also studied Ethereum official GitHub repository. Figure~\ref{fig:Githubvul} shows an example of a reported smart contract vulnerability in GitHub. 

\textbf{Data-source 3:} We used the NVD search interface to collect and extract all reported CVEs on smart contracts and their CVSS until April 2021. We searched with ``smart contracts'', ``Solidity'', and ``Ethereum''. Then, we manually extracted the reported CVEs that are related to smart contracts.    

\textbf{Data-source 4:} 
We extracted all the reported code weaknesses in SWC until April 2021. For each code weakness, we extracted the ID, title, relationships, and	test cases. 

\subsubsection{Data Cleaning and Pre-Processing}
After the initial extraction, we applied several filtering steps to obtain a clean dataset.
First, we removed posts with duplicate titles or marked as \textit{[duplicate]} in StackOverflow. 
Secondly, we manually inspected the title and the body of the question and decided if the post actually discussed smart contracts in Ethereum/Solidity.
Finally, we removed vague, ambiguous, incomplete, and overly broad posts.
As an indication for such posts, we considered the amount of negative votes and/or negative feedback.
Finally, we extracted the code of the smart contract in each post for further analysis. 

In order to pre-process the collected GitHub data, we removed the duplicates based on the description of the vulnerability. For further analysis, we also extracted the code of the smart contract that contained the vulnerability. 

We also double-checked data collected from both the NVD database and SWC registry for duplicates. 
After this stage, we had a clean dataset with records from the four data sources.

Table~\ref{table:datasetSummary} shows a summary of the collected dataset after applying the aforementioned cleaning and pre-processing steps.

\subsection{Data Analysis and Classification Categories}
\label{sec:met_categories}

To analyze and label the cleaned dataset, we manually inspected each record and read the description of each vulnerability and weakness. 
Then, to define the categories of vulnerabilities and weaknesses in smart contracts, two experts (i.e., a software engineering expert and a cybersecurity expert) together used card sorting to propose a classification scheme based on the cleaned data.
After that, we discussed each category and gave it a name based on the categories that were defined by Beizer~\citep{beizer1984software}.
To find out root causes of the categories, we analyzed the question and/or the answer information.
Finally, we followed the same categorization approach to propose a classification scheme for the impact of the recorded vulnerabilities and weaknesses.

For each record in the cleaned dataset, we created a card containing the information of the vulnerability as collected from the original data source.
Then, the two experts determined the category of the vulnerability (RQ1) independently based on its root cause. As well as the impact category (RQ3) was decided based on its consequence on the software product (i.e., smart contracts).
When cards did not fit into any of the existing categories, we extended the schemes accordingly.
We started with a random 10\% of the data, and measured inter-rater agreement after independent labeling using Cohen's Kappa coefficient.
The Kappa value between the two experts was 0.60, showing moderate agreement \citep{viera2005understanding}.
We then clarified major disagreements to reach a common decision, and continued with the remaining 90\% of the data.
We repeated the clarification discussions again after 20\% and 40\% of the posts were labeled (with Kappa values of 0.75 and 0.82 respectively).
Finally, calculations of the Kappa coefficient at 60\%, 80\%, and 100\% resulted in values of $>0.90$, indicating perfect agreement.

The same approach was followed to label the impact of each vulnerability in the cleaned dataset.
The Kappa value was also $>0.90$ in the final discussion of the impact labeling.

\subsubsection{Classification Scheme Attribute-value Design}
To design a thorough classification scheme for smart contracts' vulnerabilities, we also followed the structured approach and recommendations of \cite{seacord2005structured}.

To eliminate the problem of having vulnerabilities that fit into multiple classifications and therefore invalidate frequency data, we added attribute-value pairs in our classification for each vulnerability. This also helped us to provide an overall picture of the vulnerability.

\begin{figure}[ht]
\centering
\includegraphics[width=.5\textwidth, bb=0 0 701 550]{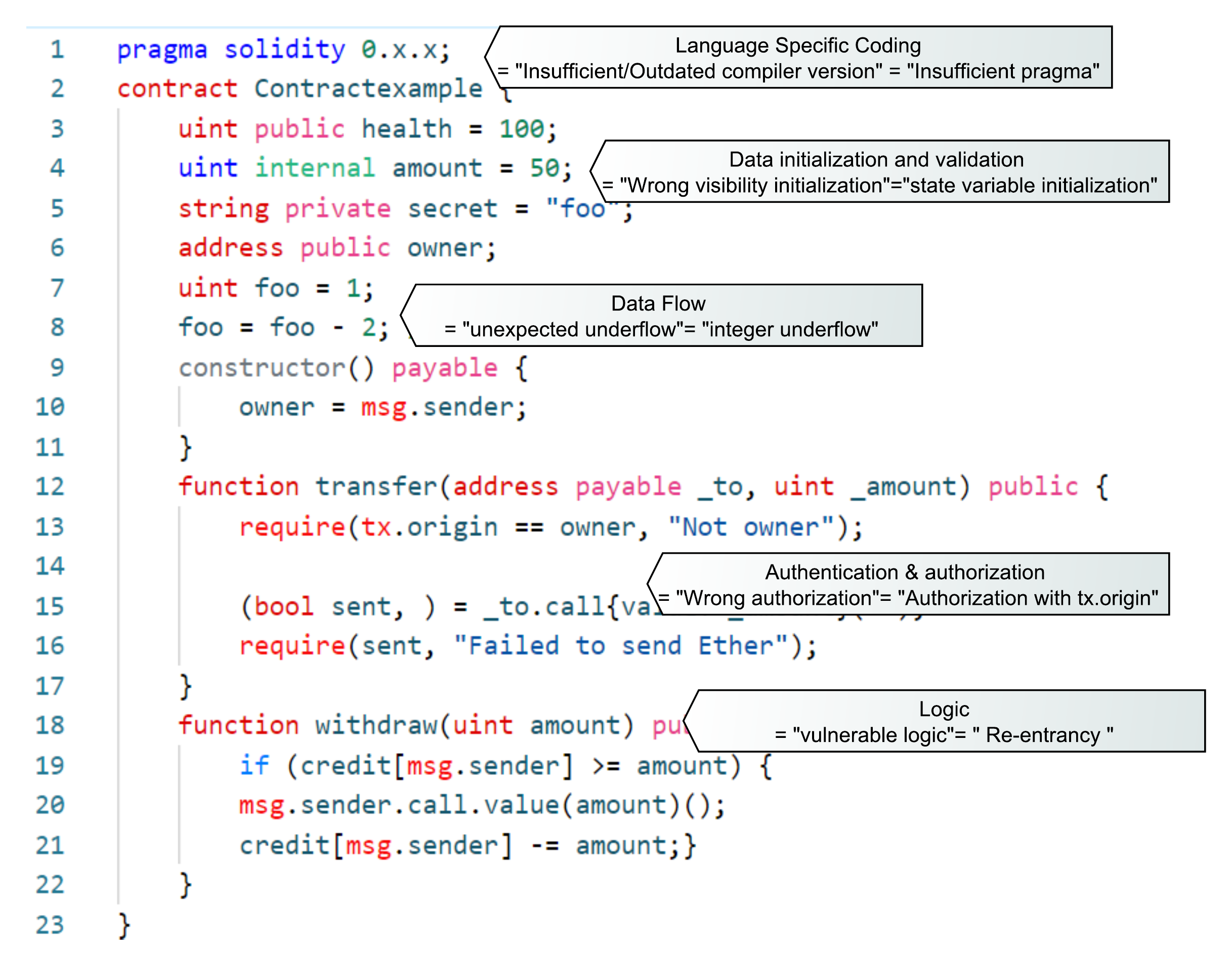}
\caption{Sample Solidity contract annotated with vulnerability attribute-value pairs}
\label{fig:exampleatt}
\end{figure}

A smart contract has many attributes such as size, complexity, performance, and other quality attributes. In this paper, we are only interested in attributes that characterize the overall vulnerability of a smart contract. Therefore, these attributes can also represent code weaknesses (that may or may not lead to vulnerabilities). All attributes and values that we list in this paper are selected based on engineering differences of the vulnerability sources of error that were concluded from expert discussions while analyzing the data (see Section~\ref{sec:met_categories}).
An example of the proposed attributes-value pairs is shown in Figure~\ref{fig:exampleatt}.

We present a sample contract (i.e., not a real contract) demonstrating solidity functions with a set of weaknesses and vulnerabilities.
In particular, the contract has multiple vulnerabilities, some of which are popular vulnerabilities, such as authorization with tx.origin and reentrancy. In addition, the contract contains another attribute that is an insufficient/ outdated compiler version, which uses insufficient pragma version 0.x.x. Another attribute is the use of wrong visibility initialization that uses wrong state variable initialization. 

The attribute-value pairs eliminate the possibility of vulnerabilities fitting into multiple classifications and therefore invalidating frequency data \cite{seacord2005structured}.

\subsection{Unifying Classification Schemes}
\label{sec:met_unification}
\begin{figure}[ht]
\centering
\includegraphics[width=.5\textwidth, bb=0 0 618 352]{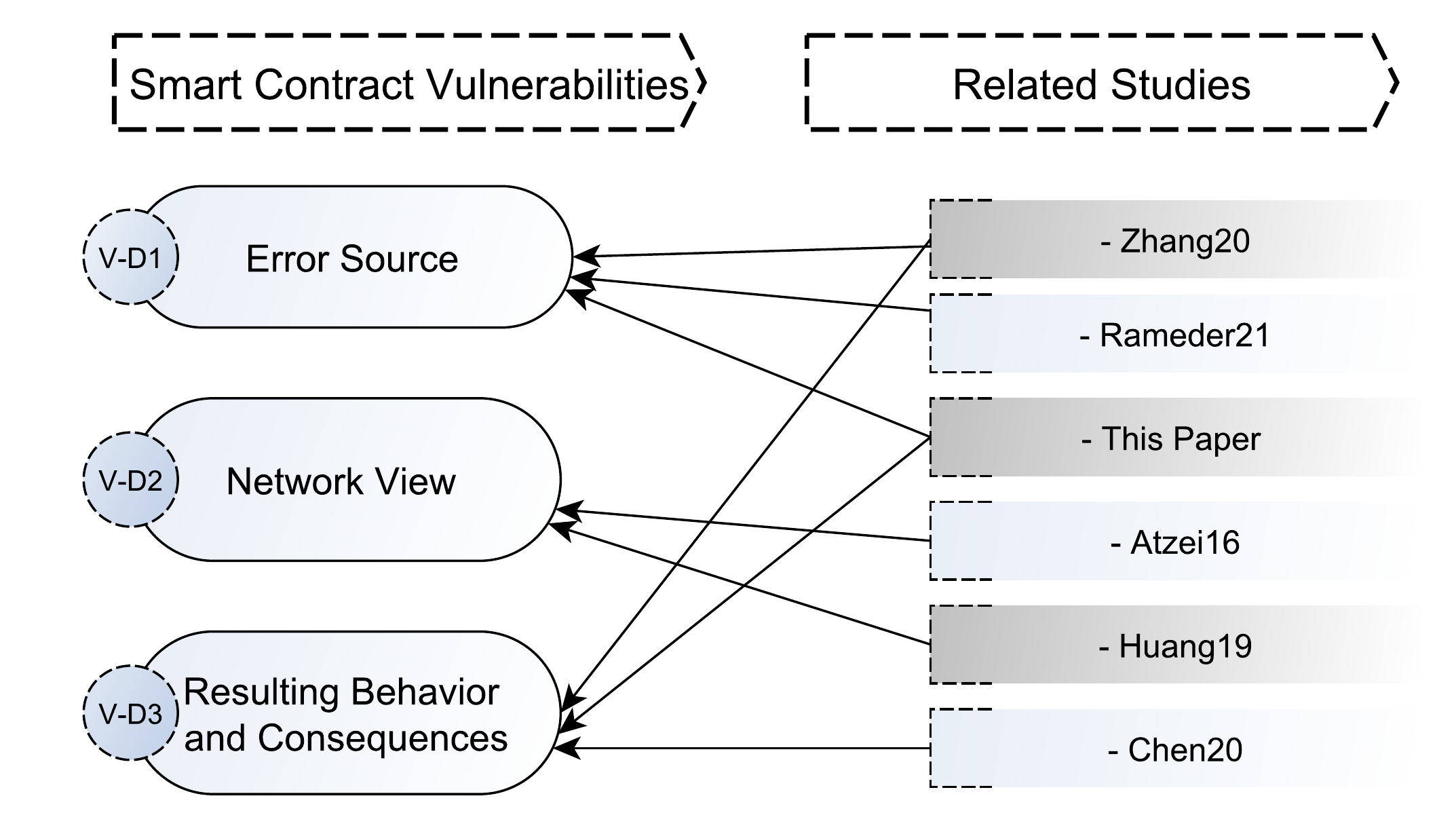}
\caption{Dimensions of Smart Contract Vulnerabilities (\emph{V-D})}
\label{fig:dimensions}
\end{figure}

To unify existing vulnerability classification schemes in the literature, we gathered all the categories proposed in the literature into an Excel sheet. The first and the second authors then discussed each category in relation to the categories proposed by us. We subsequently defined three dimensions of the problem (i.e., the vulnerability's source of error; the location of the vulnerability in the network level; the behavior and consequences arising from an exploit of the vulnerability). Afterwards, we mapped existing classifications to the defined dimensions, as illustrated in Figure~\ref{fig:dimensions}. We name these vulnerability dimensions \emph{V-D}.
The error source dimension (V-D1) describes the main cause that, when triggered, can result in executing the vulnerability, such as the logic of the contract and the data initialization. V-D2 is the network dimension, which indicates at which layer on the network the vulnerability occurs. Finally, V-D3 describes the resulting behavior and consequences of the vulnerability, which means the result of executing the vulnerability.

\begin{figure}[ht]
\centering
\includegraphics[width=.49\textwidth, bb=0 0 624 253]{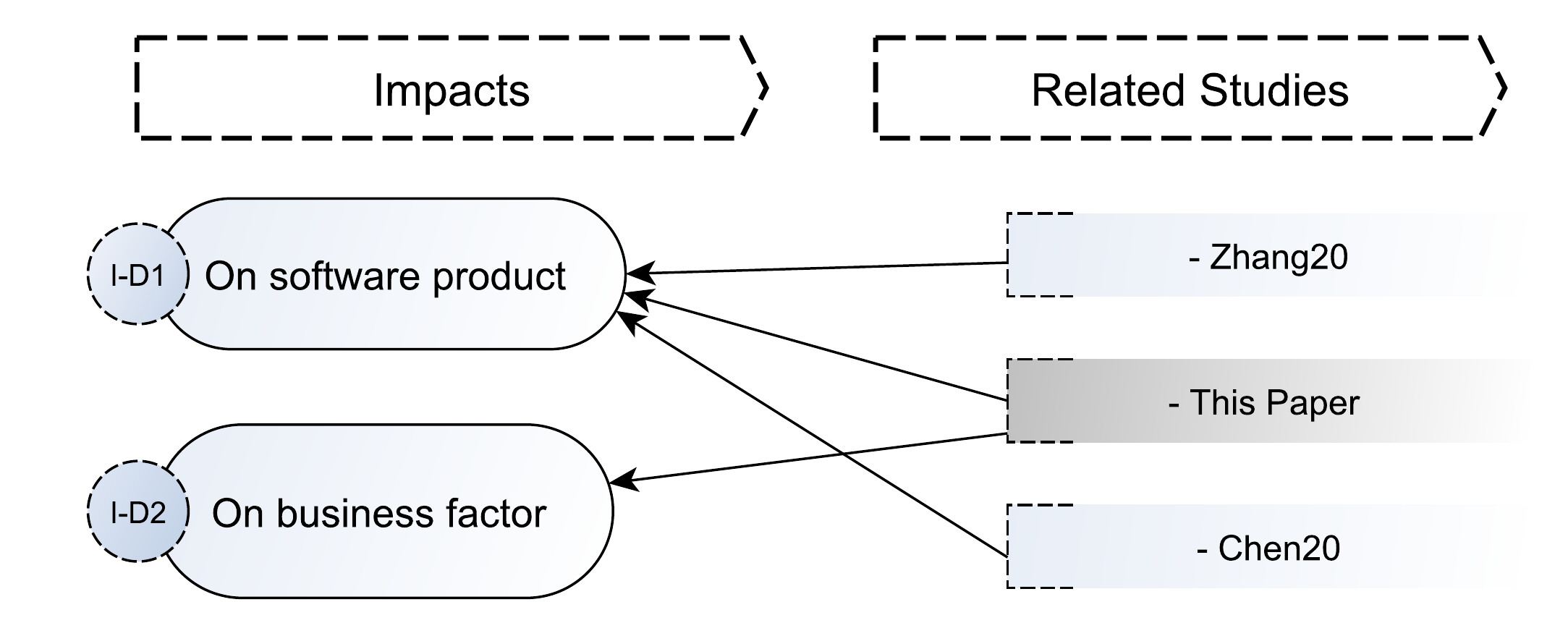}
\caption{Dimensions of Smart Contract Vulnerability Impacts (\emph{I-D})}
\label{fig:impacts}
\end{figure}

We followed the same approach to devise a thorough classification scheme for the impacts of smart contract vulnerabilities. We defined two dimensions (i.e., impact on the software product, and impact on business factors) as shown in Figure~\ref{fig:impacts}. We name these impact dimensions \emph{I-D}. I-D1 is the impact of the vulnerability on the software product itself and its resulting behavior, whereas I-D2 describes the impact of the vulnerability on the business level, e.g., losing money or important information.

\section{Results}
\label{sec:results}
In this section, we present the results of our study, and answer the proposed RQs. Our objective is to unify the existing classification schemes and define the causes, impacts, and recurrences of smart contract vulnerabilities and code weaknesses. Moreover, we propose thorough classification schemes for both the impacts and categories of smart contracts' weaknesses and vulnerabilities. 

\begin{tcolorbox}
Smart contracts' code weaknesses and vulnerabilities are found in more than half (i.e., 66.7\%) of the cleaned data from all the four data sources (Stack Overflow, GitHub, CVE and SWC). 
\end{tcolorbox}

\subsection{What categories of vulnerabilities appear in smart contracts? (RQ1)}
\label{sec:resCategories}

To answer RQ1, we followed the analysis approach in Section~\ref{sec:met_categories}, then mapped the result of our classification to other classification schemes as explained in Section~\ref{sec:met_unification}. We classified the 2143 extracted vulnerabilities and weaknesses into 47 unique smart contract weaknesses and vulnerabilities, grouped into 11 categories. Within our classification scheme, we mapped the existing literature classification schemes.

Table~\ref{table:classsch} shows a mapping between our categories and Beizer’s categories \citep{beizer1984software}. We define each category in our proposed classification scheme and show its relation to Beizer’s categories.
The categories \emph{Interface}, \emph{Dependency and upgradability}, \emph{Authentication \& authorization}, and \emph{Deployment and configurations} were added to the classification based on discussions throughout the card sorting process, and do not correspond to any category in Beizer's classification.

Table~\ref{table:mapping_vd} maps literature-based classification schemes of smart contracts vulnerabilities with our own. The categories proposed in the literature are listed in the rows, while our categories are listed in the columns of the table. The table covers all the three dimensions of \emph{V-D} discussed in Section~\ref{sec:met_unification}. As can be seen, some broad categories listed in literature essentially cover all of our classification categories.

\begin{tcolorbox}
Most literature-based classification schemes for smart contract vulnerabilities include broad categories, such as security and availability. 
\end{tcolorbox}

\begin{table*}[htbp]
\centering
\caption{Classification Scheme of Vulnerabilities in Smart Contracts}
\label{table:classsch}
\begin{tabular}{|p{3 cm}| p{13 cm} | p{0.7 cm}|} 
 \hline
Category Name &Description &Short 
\\ 
 \hline\hline
Language Specific Coding &Syntax mistakes in implementing Solidity contracts that are not captured by the Solidity compiler can introduce unexpected behavior or damage the contract. Similar to \emph{Implementation} in \cite{beizer1984software}, but specific to Solidity. & CV 
\\
\hline
Data Initialization and Validation & 
The input data type to the contract or the fields' data type in the contract are not initialized correctly. Also includes the data passed from/sent to other contracts. Corresponding to \emph{Data definition} and \emph{Data access} in \cite{beizer1984software}. 
& DV 
\\

& \textbf{Predictable resources} is a subcategory of data initialization and validation. Weaknesses and vulnerabilities resulting from using expected values in state variables or functions. These vulnerabilities can allow malicious minors to take advantage of the vulnerabilities and control the contract.
  & DV-PR\\ 
\hline
Structural Sequence \& Control & Problem with the structure of the contract control flow. Specifically, a result of incorrect control flow structure such as require, if-else conditions, assert, and loop structures. Corresponding to \emph{Flow control and sequence} in \cite{beizer1984software}. & SV-SC \\
\hline
Structural Data Flow &Problems with the structure of the contract data flow. The main difference to \emph{Data} is that \emph{Data} originates in the data fields and input parameters to the subcontracts or the contract methods. Instead, in this category, changing data fields in a wrong way during and after the execution of the contract leads to issues. Corresponds to \emph{Data-flow anomaly} in \cite{beizer1984software}. & SV-DF \\
\hline
Logic & Inconsistency between the intention of the programmer and the coded contract, and not one of the other categories. Corresponds to \emph{Logic} in \cite{beizer1984software}.
& LV\\ 
\hline
Timing \& Optimization & Performance and timing issues that can affect execution time and results in abnormal responsiveness under a normal workload. Corresponds to \emph{Performance and timing} in \cite{ beizer1984software}, considering the existence of the Ether/gas concept. & TV \\%&\cite{rameder2021systematic}, \cite{zhang2020framework},\cite{chen2020defining},\cite{alharby2018blockchain}, \cite{khan2020survey} \\
\hline
Compatibility & Required software and packages are not compatible with the available resources (e.g., operating system, CPU architecture). Corresponds to \emph{Configuration sensitivity} in \cite{beizer1984software}& CoV \\%&\cite{zhang2020framework}\\
\hline
Deployment \& Configurations & Weaknesses during deployment of implemented contracts on the Ethereum blockchain. & DL\\%& \cite{rameder2021systematic},\cite{chen2020survey}\\
\hline
Authentication \& Authorization & Vulnerabilities allow malicious actors to take control over the contract.
& SV\\%&\cite{rameder2021systematic}
\hline
Dependency \& Upgradability & Upgrades in a smart contract breaking the dependencies in the new contract. 
 & UV
 \\
\hline
Interface & Vulnerabilities and weaknesses in the interface of smart contracts, e.g., when the contract is functioning correctly, but the interface is showing a wrong output that contradicts the contract’s execution logs and transaction logs. & IB
\\

\hline
\end{tabular}
\end{table*}

\begin{table*}[htbp]
\centering
\caption{Mapping Literature-based Classifications to V-D. The rows marked \emph{V-D1} refer to the source of error dimension, rows marked \emph{V-D2} refer to the network view dimension, and rows marked \emph{V-D3} refer to the resulting behavior and consequences dimension. {$\subset$}* indicates that the category in the V-D classification is a subset of the corresponding category in the literature marked by *. *{$\subset$} means the category in the literature is a subset of a proposed category in V-D. Finally, {=} means the categories are identical.}
\label{table:mapping_vd}
\begin{tabular}{|l||*{10}{c|}}\hline
\backslashbox{Literature}{Ours}

&\makebox[1.5em]{CV}&\makebox[1.5em]{DV}&\makebox[1em]{SV}
&\makebox[1em]{LV}&\makebox[1.2em]{TV}&\makebox[2em]{CoV}&\makebox[1.2em]{DL}&\makebox[1.2em]{SV}&\makebox[1.2em]{UV}&\makebox[1.2em]{IB}\\\hline\hline
\tikzmark[xshift=-8pt,yshift=1ex]{m}Codifying \citep{alharby2017blockchain}&\textbf{$\subset$}*&&&\textbf{$\subset$}*&&&&&&\\\hline 
Data* \citep{zhang2020framework} &*{$\subset$}&*{$\subset$}&&&&&&&&\\\hline
Description* \citep{zhang2020framework} &*{$\subset$}&&&&&&&&&\\\hline
Interaction* \citep{zhang2020framework} &*{$\subset$}&*{$\subset$}&&*{$\subset$}&&&&&&\\\hline
Interface* \citep{zhang2020framework} &&&&&&&&&&=\\\hline
Logic* \citep{zhang2020framework} &&&&=&&&&&&\\\hline
Standard* \citep{zhang2020framework} &*{$\subset$}&&&&&&&&&\\\hline
Authentication* \citep{rameder2021systematic} &&&&&&&&=&&\\\hline
Arithmetic* \citep{rameder2021systematic} &&&&=&&&&&&\\\hline
Bad Coding Quality* \citep{rameder2021systematic} &=&&&&&&&&&\\\hline
Environment Configuration* \citep{rameder2021systematic} &&&&&&&=&&&\\\hline
 \tikzmark[xshift=-8pt,yshift=-1ex]{l}Deprecated* \citep{rameder2021systematic} &*{$\subset$}&&&&&&&&&\\\hline \hline

 \tikzmark[xshift=-8pt,yshift=1ex]{x}Solidity* \citep{atzei2016survey}& *\textbf{$\subset$} & \textbf{$\subset$}* &\textbf{$\subset$}*& \textbf{$\subset$}* &\textbf{$\subset$}* & & \textbf{$\subset$}*& \textbf{$\subset$}*&\textbf{$\subset$}*&\tikzmark[xshift=3.5em]{a} \\ \hline 
EVM* \citep{atzei2016survey}&&&&&&\textbf{$\subset$}*&&&&\\\hline
\tikzmark[xshift=-8pt,yshift=-1ex]{y}Blockchain* \citep{atzei2016survey} &&&&&&&\textbf{$\subset$}*&&& \tikzmark[xshift=3.5em]{b} \\ \hline \hline 

 \tikzmark[xshift=-8pt,yshift=1ex]{w}Security* \citep{alharby2017blockchain}&\textbf{$\subset$}*&\textbf{$\subset$}*&\textbf{$\subset$}*&\textbf{$\subset$}*&\textbf{$\subset$}*&\textbf{$\subset$}*&\textbf{$\subset$}*&\textbf{$\subset$}*&\textbf{$\subset$}*&\textbf{$\subset$}*\\\hline
Privacy* \citep{alharby2017blockchain}&{$\subset$}*&{$\subset$}*&&{$\subset$}*&&&&&&\\\hline
Performance* \citep{alharby2017blockchain}&&&&&=&&&&&\\\hline
Availability* \citep{chen2020defining}&{$\subset$}*& {$\subset$}*&{$\subset$}*&{$\subset$}*&{$\subset$}*&{$\subset$}*&{$\subset$}*&{$\subset$}*&{$\subset$}*&{$\subset$}*\\\hline
Maintainability* \citep{chen2020defining} &{$\subset$}*&&{$\subset$}*&{$\subset$}*&&{$\subset$}*&&&{$\subset$}*&\\\hline
 \tikzmark[xshift=-8pt,yshift=-1ex]{z}Reusability* \citep{chen2020defining}&{$\subset$}*&{$\subset$}*&{$\subset$}*&{$\subset$}*&&&{$\subset$}*&{$\subset$}*&&{$\subset$}*\\\hline

\end{tabular}
\drawbrace[brace mirrored, thick]{x}{y}
\drawbrace[brace mirrored, thick]{w}{z}
\drawbrace[brace mirrored, thick]{m}{l}
\annote[left]{brace-1}{V-D2}
\annote[left]{brace-2}{V-D3}
\annote[left]{brace-3}{V-D1}
\end{table*}

The following subsections present the key findings related to the defined categories. For each category, we define and explain the most critical vulnerabilities and weaknesses as agreed by the two raters. 
We give examples for some vulnerabilities and weaknesses, which are directly taken from our dataset.
For the full list of vulnerability and code weakness definitions, we refer to our published dataset~\citep{dataset}.

\subsubsection{Language Specific Coding Vulnerabilities and Weaknesses}
\begin{table}[htbp]
\centering
\caption{Language Specific Coding Attribute}
\label{table:CVattribute}
\begin{tabular}{||p{2 cm} |p{6 cm}||} 
 \hline
 Attribute & Values\\ [0.5ex] 
 \hline
 Language specific coding & pragma version, fallback function, pre-defined functions in the language, language standards, language defined libraries, syntax issues (not detectable by the compiler), style guide and recommended language patterns, experimental language features, deprecated code, unsafe language features. \\
\hline
\end{tabular}
\end{table}

In this category, smart contracts' vulnerabilities and weaknesses result from language-based errors not captured by the compiler. The source of error of this category can be in the language pre-defined functions, events, libraries, and/or language standards. Table~\ref{table:CVattribute} shows the attribute-value pair for this category.

We define 14 Language Specific Coding vulnerabilities and weaknesses that can result in undesirable state in a smart contract or can be used by attackers in their favor. Next, we show a sample of these vulnerabilities and weaknesses. 

\renewcommand{\labelitemi}{$\blacksquare$}
 \renewcommand\labelitemii{$\square$}

\textbf{Insufficient compiler version or pragma version --- CV\#1.} A so-called version \emph{pragma} should be included in the source code of smart contract to reject compiling the contract using incompatible compiler versions. When using a version \emph{pragma} in the contract which is later than the selected compiler, this may introduce incompatible changes and lead to vulnerabilities in compiled smart contract code. Moreover, future compiler versions may handle language constructs in a way that introduces unclear changes affecting the behavior of the contract as shown in Listing~\ref{lst:cv1}. 

\begin{lstlisting}[language=Python, caption=Version pragma, label={lst:cv1}]
pragma solidity ^0.6.3; // weakness/ vulnerability 
pragma solidity 0.6.3; 
\end{lstlisting}

\textbf{Fallback function not payable --- CV\#2. }Smart contracts written in Solidity versions 0.6.0+ should have the fallback function split up into a \emph{receive()} and a \emph{fallback()} function (i.e., a new fallback function that is defined using the \emph{fallback} keyword and a \emph{receive} ether function defined using the \emph{receive} keyword). If present, the \emph{receive} function is called whenever no parameters are provided in the call. The \emph{receive} function is implicitly payable. The new \emph{fallback} function is called when no other function matches. However, if \emph{fallback()} is not payable and \emph{receive()} does not exist, transactions not matching any other function which send ether will revert and result in an undesirable state in the contract.

\textbf{Fallback function does not exist --- CV\#3.} 
In addition to CV\#2, when sending Ether from a contract to another contract without calling any of the receiving contract’s functions, sending the Ether will fail if the receiver contract has no fallback function. Thus, a payable fallback function should be added to the receiver before deployment. Otherwise, there is no way to receive the Ether unless the sender has previous knowledge of the exact functions of the receiving contract, which is not usually the case.

\textbf{Violating splitting Fallback function --- CV\#4.} Smart contracts written in Solidity versions 0.6.0+ should have the fallback function split up to \emph{receive()} and \emph{fallback()} (i.e., a new Fallback function that is defined using the \emph{fallback} keyword and a \emph{receive} ether function defined using the \emph{receive} keyword). If present, the \emph{receive} function is called whenever the call data is empty. The \emph{receive} function is implicitly payable. The new \emph{fallback} function is called when no other function matches. The fallback function in Listing~\ref{lst:cv4} can be payable or not, however, if it is not payable then transactions not matching any other function which send value will revert.
\begin{lstlisting}[language=Python, caption=Fallback function, label={lst:cv4}]
contract payment{
  mapping(address => uint) _balance;
  fallback() payable external {
    _balance[msg.sender] += msg.value;
  }
}\end{lstlisting}

 \textbf{Violating modifier definition --- CV\#5.}
Solidity provides modifiers that are used to change the behavior of functions in a declarative way. They can be used to enforce pre/post-conditions on the execution of a function. The \_ operator should be used in defining a modifier, and starting from Solidity version 0.4.0+ a semicolon should be added after the \_ operator. The operator represents the actual code of the function that is being modified. Thus, the code for the function being modified is inserted where the \_ is placed in the modifier. Missing the \_ operator might generate unwanted results. For example, in Listing~\ref{lst:modifierDef}, line 8, every time transferOwnership is invoked, the onlyOwner modifier will get into play first. If the owner invokes it, then the control flow will reach the \_ operator, so the transferOwnership statements will be executed. Otherwise, the execution will just throw, revert, and exit.
 
\begin{lstlisting}[language=Python,label={lst:modifierDef}, caption=Violating \emph{modifier} definition]
contract owned {
  address public owner;
  function owned() {
    owner = msg.sender;}
  modifier onlyOwner {
    if (msg.sender != owner) throw;
    _;  }
  function transferOwnership(address newOwner) onlyOwner {
    owner = newOwner; }}\end{lstlisting}

\textbf{Manipulated language standard --- CV\#6}
Ethereum has adopted many standards to guarantee the composability of smart contracts and Dapps. Those standards are in Ethereum's official EIPs and include token \footnote{An Ethereum token can represent anything, including lottery tickets, financial assets, a fiat currency like USD, an ounce of gold, etc.} standards. For example, ERC-20 is a token technical standard that allows developers to implement tokens of cryptocurrencies. It contains nine unique functions and two events to guarantee the possibility of exchanging tokens based on ERC-20 with other ERC-20 tokens. Any modification on the function name, parameter types, and the return value in the standard might change its functionality and leave the developer believing it is the same as ERC-20. The implementation of ERC-20 in any contract shall strictly be the same as in the standard template.

 \textbf{Violating call-stack depth limit --- CV\#7.}
In Ethereum, the call-stack has a hard limit of 1024 frames. Each time the contract calls an external contract, the call-stack depth of transaction increases by one. Thus, when the number of calls exceeds the limit of the call-stack, an exception is thrown and the call is aborted by Solidity. Moreover, Solidity does not support exceptions in low-level external calls. Therefore, a malicious actor can recursively call a contract 1023 times, then call a victim contract to reach the call-stack depth limit. This will fail any subsequent call made by the victim contract without the victim contract owner being aware of the attack. Recently, EIP 150\footnote{https://github.com/ethereum/EIPs/blob/master/EIPS/eip-150.md} makes it impossible to reach stack depths of 1024, effectively eliminating call depth attacks

 \textbf{Insufficient Address split --- CV\#8.} Starting from Solidity 0.5.0+, the address should be split to address and address payable, where only address payable provides the transfer function. Otherwise, sending tokens to ``unpayable'' addresses will be reverted. Moreover, there is no way to convert an address to address payable.

\textbf{Mixing \emph{pure} and \emph{view} --- CV\#9.}
In Solidity, \emph{pure} and \emph{view} are function modifiers that describe how the logic in that function will interact with the contract's state. Functions that are declared \emph{view} promise not to modify the state, while functions that are declared \emph{pure} promise not to read or write the state.By using no specifier, the state can be read as well as modified. Developers can mix these two modifiers by replacing \emph{view} with \emph{pure} or missing any of these modifiers, resulting in unexpected state changes or incorrect reads from the state . 

 \textbf{Using of \emph{balance} as attribute to the contract --- CV\#10}
One of the features of Solidity is that contracts inherit all members from \emph{Address}, meaning that the keyword \emph{this} is the pointer to the current instance of the type derived from Address. In other words, if the developer wants to access members of the address type (e.g. the balance) of the current contract instance, then the developer should use \emph{this} and should use \emph{balance} as an attribute of the address type, not the contract as shown in Listing~\ref{lst:cv11}. We noticed a confusion in using balance and other address attributes as if they are attributes of the contract. 
\begin{lstlisting}[language=Python, caption=Using of \emph{balance} as attribute to the contract , label={lst:cv11}]
function getSummary() public view returns(
  uint, uint, uint, uint, address
){
  return (
    minimumContribution,
     this.balance, //incorrect
    address(this).balance,// correct 
    requests.length,
    approversCount,
    manager
  );
}\end{lstlisting}

\textbf{Unsafe delegatecall (code injection)--- CV\#11 }
A special variant of a message call in Solidity is \emph{delegatecall}. With this feature, the contract can be executed in the callee's context, while msg.sender and msg.value remain unchanged. Consequently, it allows an external contract to modify the storage of the calling contract. This can be exploited by a malicious caller to manipulate the caller's contract state variables and take full control over the balance.

\textbf{Variable shadowing --- CV\#12}
Solidity supports ambiguous naming when inheritance is used. For instance, contract Alpha with a variable V could inherit contract Beta that also has a state variable V defined. Consequently, there would be two versions of V, one accessed from contract Alpha and the other from contract Beta. In complex contract systems, this condition might go undetected and ultimately cause security issues. Also, this can also occur at the contract level (e.g., a contract with more than one definition at the contract and function level). 

\textbf{Deprecated Solidity code --- CV\#13}
As Solidity evolves, several of its functions and operators are deprecated. Making use of them leads to poor code quality. It is strongly discouraged to use deprecated Solidity language code with new major versions of the Solidity compiler, since it can cause unwanted behavior and vulnerabilities. 

\textbf{Experimental Language Features --- CV\#14}
Similar to CV \#13, it is strongly discouraged to use experimental Solidity language features since it can cause undesired behavior and code weaknesses.

\subsubsection{Data Vulnerabilities and Weaknesses}
Most of the data vulnerabilities result from the use of wrong or insufficient data types, or passing wrong data formats to arguments without knowing the exact required type. Moreover, organizing the memory and storage in Solidity is the responsibility of programmers, which many developers are not used to do. Table~\ref{table:dataattribute} shows the attribute-value pairs for this category.
\begin{table}[htbp]
\centering
\caption{Data vulnerabilities and weaknesses Attribute}
\label{table:dataattribute}
\begin{tabular}{||p{2 cm} |p{6 cm}||} 
 \hline
 Attribute & Values\\ [0.5ex] 
 \hline
 Data vulnerabilities and weaknesses & Insufficient/wrong data type, wrong addresses initialization, writing on arbitrary locations, insufficient memory and storage management, improper data validation, improper state variable initialization, data pointer initialization, function pointer initialization. \\
\hline
\end{tabular}
\end{table}

\textbf{Violating explicit data location --- DV\#1.} For Solidity versions 0.5.0+, an explicit data location for all variables of type struct, array, or mapping is mandatory. This also applies to function parameters and return variables. For instance, \emph{calldata} is a special data location that contains the function arguments, which is only available for external function call parameters. If \emph{calldata} is not included in the initialization, it results in unexpected values as shown in Listing~\ref{lst:dv1}.

\begin{lstlisting}[language=Python, label={lst:dv1}, caption=Using of \emph{torage} instead of \emph{memory}]
contract StructExample {

  struct SomeStruct {
    int someNumber;
    string someString;
  }
  SomeStruct[] someStructs;

  function addSomeStruct() {
    SomeStruct storage someStruct = SomeStruct(123, "test");// insufficient use
    SomeStruct memory someStruct = SomeStruct(123, "test");// correct
    someStructs.push(someStruct);
  }
}
\end{lstlisting}

\textbf{Using of \emph{storage} instead of \emph{memory} --- DV\#2.} In addition to \emph{calldata}, Solidity provides two more reference types to comprise structs, arrays, and mappings called \emph{storage} and \emph{memory}. The Solidity contract can use any amount of memory (based on the amount of Ether that the contract owns and can pay for) during execution. However, when execution stops, the entire content of the memory is wiped, and the next execution will start fresh. The \emph{storage} is persisted into the blockchain itself, so the next time the contract executes, it has access to all the data it previously stored in its storage area. Confusing storage and memory can result in data loss.

\textbf{Violating array indexing --- DV\#3.} Developers are making numerous mistakes when initializing and accessing arrays in Solidity. Most of the time, discovering these violations is not easy, especially if there is no syntax error or an error that can be detected by the compiler. This can result in returning incorrect values. A clear violation of array indexing is shown in Listing~\ref{lst:arrayViol}, line 9, where the developer is trying to access a single element in a 3-dimensional array, but only provides two sets of square brackets. Therefore, the developer is returning an array instead of a single Object.

\begin{lstlisting}[language=Python,label={lst:arrayViol}, caption=Violating arrays indexing ]
contract Game {
  struct User{
    address owner; }
User[][10][10] public gameBoard;
User memory mover = gameBoard[_fromX][_fromY][0];
   function addUser (uint _x, uint _y) public {
        gameBoard[_x][_y].push(User(msg.sender, 10, 5, 5, 5, 5));}
        function moveUser (uint _fromX, uint _fromY) public {
        User memory mover = gameBoard[_fromX][_fromY]; //incorrect access
        if (mover.owner != msg.sender)return;}} 
\end{lstlisting}

\textbf{Hard-coded address --- DV\#4}
An existing bad practice is the use of hard-coded addresses in smart contract code, as shown in Listing~\ref{lst:dv4}. Any incorrect or missing digit in the address may result in losing Ether, in case Ether is sent to that wrong address, or in unexpected outcomes. This vulnerability is also known as ``Transfer to orphan address'' \cite{atzei2016survey}. 

\begin{lstlisting}[language=Python,label={lst:dv4}, caption=Hardcoded address ]
address recieveraddress ;
function initializeAddress1 () {
  recieveraddress = 0x98081c...8e5ace; //hardcoded address}\end{lstlisting}
\textbf{Improper data validation --- DV\#5}
It is necessary to validate input from untrusted sources, such as external libraries or contracts before integrating it into any contract logic. 

\textbf{Unintentional Write to arbitrary storage location --- DV\#6}
Because Solidity storage is not dynamically allocated, it can lead to unpredictable behavior, unauthorized data access, and other vulnerabilities, especially if the data location of data types like structs, mappings, and arrays is not clarified and allowed to overwrite entries of other data structures.

\subsubsection{Predictable Data Values and Resources}
We encountered several issues in Solidity smart contracts that relate to values that can be guessed even though they are intended to serve as an element of randomness.

\textbf{Timestamp dependency --- DV-PR\# 7}
To keep contracts safe from malicious actors, developers should avoid using the block variables as a source of randomness or as part of triggering conditions for executing significant operations in their contracts, such as transferring Ether. When submitting blocks, miners determine the value of block variables such as block.timestamp, block.coinbase, and block.difficulty. Thus, these values can affect the contract's outcome and can be used to benefit the attacker. For example, Listing~\ref{lst:dv-pr7} shows an insecure lottery contract in which block.timestamp is used as a source of entropy.

\begin{lstlisting}[language=Python,label={lst:dv-pr7}, caption=Insecure lottery contract using block variables]
function setWinner() public {

  bytes32 hash = keccak256(abi.encode(block.timestamp));
  bytes4[2] memory x0 = [bytes4(0), 0];
  assembly {
    mstore(x0, hash)
    mstore(add(x0, 4), hash)
  } \end{lstlisting}

\textbf{Blockhash dependency --- DV-PR\# 8}
Using blockhash has the same risks as block.timestamp in DV-PR \#7,  especially when used in critical operations such as Ether transfer. It can lead to serious attacks as malicious miners can tamper with the blockhash and take full control over the contract. 

\textbf{Bad random number generation--- DV-PR\# 9}
Using random numbers is not avoidable in some smart contracts, e.g., games or lotteries. It is important that the randomness is not based on global blockchain variables, as that leaves the contract open to manipulation by malicious miners similar to DV-PR\#7 and DV-PR\#8.

\subsubsection{Sequence and Control Vulnerabilities and Weaknesses}
This category of vulnerabilities is corresponding to incorrect control structure and loop control statements. These can be exploited and help the attacker to steal money in the contract. Moreover, they can also result in losing all the money in the contract without even being attacked, just because of vulnerabilities in these structures. 
 Table~\ref{table:seqattribute} shows the attribute-value pairs for this category.

\begin{table}[htbp]
\centering
\caption{Sequence and control vulnerabilities and weaknesses}
\label{table:seqattribute}
\begin{tabular}{||p{2 cm} |p{6 cm}||} 
 \hline
 Attribute & Values\\ [0.5ex] 
 \hline
 Sequence and control & Wrong use of assert, wrong use of require. \\
\hline
\end{tabular}
\end{table}

 \textbf{Using assert instead of require --- SV-SC\#1.} The \emph{assert} statement should be only used for conditions that indicate you have an internal vulnerability in the contract code. The \emph{require} statement should be used to check valid conditions (e.g., state variables conditions are met, validate input, and validate return value from external contracts). A valid code with correct functions should never fail \emph{assert} conditions. Otherwise, there is a vulnerability in the contract and something unexpected has happened.
 In smart contracts, \emph{assert} can consume all the gas in the contract as shown in Listing~\ref{lst:sv-sc1}. If the example is tested with run(8), the function runs successfully and 1860 gas will be consumed based on the cost of the function and the loop iterations\footnote{The actual gas costs are stated in the Solidity documentation and depend on numerous factors, such as the executed functions and the used data types.}. If it is tested with run(15), then the function passes assert, fails at require and the first loop only will be executed and consume 1049 gas. Finally, testing the same example with run(25) causes the function to fail the assert statement. The execution continues and thus iterates 25 times through the loop, resulting in a very high cost of gas.
 
 \begin{lstlisting}[language=Python, caption=Using assert instead of require, label={lst:sv-sc1}]
contract Test {
    function run(uint8 i) public pure {
        uint8 total = 0;
        for (uint8 j = 0; j < 10; j++)
          total += j;
        assert (i < 20);
        require (i < 10);
        for (j = 0; j < 10; j++)
          total += j;
    }\end{lstlisting}

\subsubsection{Data Flow Vulnerabilities and Weaknesses}
These are vulnerabilities and weaknesses in the data flow of smart contracts, so that data fields are changing unexpectedly or incorrectly. We defined two vulnerabilities that belong to this category. 
Table~\ref{table:seqdataflowattribute} shows the attribute-value pairs for this category.

\begin{table}[htbp]
\centering
\caption{Data flow vulnerabilities and weaknesses}
\label{table:seqdataflowattribute}
\begin{tabular}{||p{2 cm} |p{6 cm}||} 
 \hline
 Attribute & Values\\ [0.5ex] 
 \hline
 Data flow & Unexpected integer overflow/underflow, unexpected conversion in data values, unexpected arithmetic operation behavior 
 \\
\hline
\end{tabular}
\end{table}

 \textbf{Updating storage in fallback functions--- SV-DF\#1.} Upon receiving Ether without a function being called, either the receive Ether or the fallback function is executed. If the contract does neither have a receive nor a fallback function, the Ether will be rejected by throwing an exception. During the execution of one of these functions, the contract can only rely on the passed gas (i.e., 2300 gas) being available to it at that time. This stipend is not enough to modify storage. However, we found that developers sometimes are updating state variables, trying to write to the storage in the fallback functions. Updating the variables with such a gas limit will fail.
 If the data flow of the contract depends on the failed state variables, this results in an incorrect data flow and in unexpected outcomes. 

 \textbf{Arithmetic operation/calculation overflow/ underflow --- SV-DF\#2.} An overflow can happen as a result of an arithmetic operation or calculation that falls outside the range of a Solidity data type, resulting in unwanted behavior or unauthorized manipulation of the contract balance. Underflow happens when an arithmetic operation reaches the minimum size of a type. This is a data flow vulnerability, as the code of the contract does not perform correct validation on the numeric inputs and the calculations. In addition, the Solidity compiler does not enforce detecting integer overflow/underflow. 
 An example is shown in Listing~\ref{lst:overflow}, where the computation of \emph{mask} overflows at x $>$= 248.
\begin{lstlisting}[language=Python,label={lst:overflow}, caption=Arithmetic Operation/calculation overflow ]
uint256 public MAXUINT256 = 2*256 - 1;
for (uint256 x = 0; x < 255; x++) {
   var mask = MAXUINT256 * (2 ** x);
   var key = signature & bytes32(mask);}\end{lstlisting}
   
   \subsubsection{Logic Vulnerabilities and Weaknesses}
This category reflects inconsistencies with the contract and the programmer’s intention, which is usually mentioned in the question information. These issues relate to a number of reasons, e.g., vague developer intentions, misunderstanding of language components, incorrect usage of Math, and incorrect gas predictions.

\textbf{Greedy contract--- LV\#1.}
This vulnerability occurs when implementing a contract logic that is only locking Ether balance all the time because of its inability to access the external library contract to transfer Ether. For instance, the contract logic may only accept transferring money based on a specific value in the code, which happens to be unreachable due to incorrect logic. In this case, the Ether will be locked in the deployed contract forever. In the example of Listing~\ref{lst:greedy}, the function \emph{refundMoney()}, line 8, does not decrease the \emph{weiRaised} value, meaning that once starting a refund, the developer can no longer use \emph{forwardAllRaisedFunds()} to drain the contract. This code weakness would be triggered even in the regular course of action and Ether in this contract is stuck. It can receive any funds but the received funds can never be retrieved.

\begin{lstlisting}[language=Java, label={lst:greedy}, caption=Greedy Contract]
contract SwordCrowdsale is Ownable {
    //amount of raised money in wei
    uint256 public weiRaised;
    bool public isSoftCapHit = false;
    //send ether to the fund collection wallet
function forwardAllRaisedFunds() internal {
    wallet.transfer(weiRaised);}
function refundMoney(address _address) onlyOwner public {
    uint amount = contributorList[_address].contributionAmount;
    if (amount > 0 && _address.send(amount)) { 
    //user got money back
    contributorList[_address].contributionAmount =  0;
    contributorList[_address].tokensIssued =  0;}
    }\end{lstlisting}
\textbf{Transaction order dependency --- LV\#2}  
The vulnerability arises when a contract's logic is dependent on the order in which transactions are executed and processed in a block. It is a type of race condition inherent to Blockchains. By manipulating the order of transaction processing, malicious miners can take advantage of the contract and benefit from it. Therefore, the logic of the contract should not rely on the transaction order. 

\textbf{Call to the unknown --- LV\#3}
This vulnerability arises when a function unexpectedly invokes the fallback function of the recipient. Consequently, malicious code can be introduced. For example, the unknown call could trigger the recipient's fallback function, allowing malicious code to execute.Also, this can be done via direct call, delegatcall, send, or only call functions. In the MultiSig Wallet Attack\footnote{https://blog.openzeppelin.com/on-the-parity-wallet-multisig-hack-405a8c12e8f7/}, an attacker exploited this vulnerability to steal 30M USD from the Parity Wallet. Another example is shown in Listing~\ref{lst:Unknown}. In which, pong function uses a direct call to invoke Alice’s ping. However, if the interface of contract Alice by mistake was mistyped by declaring the parameter as \emph{int} instead of \emph{unit} and Alice has no function with \emph{int} type, then the call to ping results in calling  Alice’s fallback function.

\begin{lstlisting}[language=Python, ,label={lst:Unknown},caption=Call to the unknown \cite{atzei2016survey}]
contract Alice {function ping(uint) returns (uint)}
contract Bob {function pong(Alice c){c.ping(42);}}   
\end{lstlisting}

\textbf{DoS by external contract --- LV\#4}
It is possible for external calls to fail and throw exceptions or revert the transaction. Inefficient management of these calls in the contract logic can lead to critical vulnerabilities, such as a Denial of Service (DoS) or loss of funds.

\subsubsection{Timing and Optimization Vulnerabilities and Weaknesses}
Performance and timing vulnerabilities/weaknesses in smart contracts usually affect the gas amount in the contracts. In the following, we define 2 vulnerabilities belonging to this category.

\textbf{Unbounded loops --- TO\#1.} In Solidity, iterating through a potentially unbounded array of items can be costly, as exemplified in \emph{getNotes()} in Listing~\ref{lst:unbounded}. 
Since the array \emph{notes} is provided as an input, the smart contract has no control over the maximum length, allowing a malicious actor to send in large arrays.

\textbf{Creating subcontracts cost --- TO\#2.} Contract deployments are very expensive operations. For instance, deploying a contract for every patient in Listing~\ref{lst:unbounded} is very costly. A malicious developer can use this weakness to cost the owner of the contract more Ether.   

\begin{lstlisting}[language=Python, ,label={lst:unbounded},caption=Unbounded loops and creating subcontracts]
contract MedicalRecord {
struct Doctor {
    bytes32 name;
    uint id;}
struct Note {
    bytes32 title;
    bytes32 note;}
function getNotes()
    view
    public
    isCurrentDoctor
    returns (bytes32[], bytes32[])
{
    bytes32[] memory titles = new bytes32[](notes.length);
    bytes32[] memory noteTexts = new bytes32[](notes.length);
    for (uint i = 0; i < notes.length; i++) {
        Note storage snote = notes[i];
        titles[i] = snote.title;
        noteTexts[i] = snote.note;
    }
    return (titles, noteTexts);}
\end{lstlisting}

\textbf{Costly state variable data type --- TO\#3}
Because of the padding rules, the byte[] data type consumes more gas than a byte array. In addition, declaring variables without constant consumes more gas than one declared with a constant. This weakness is also reported in \cite{chen2020defining}.

\textbf{Costly function type --- TO\#4}
A function declared public rather than external and not utilized within the contract consumes more gas on deployment than it should.This weakness is also reported in \cite{chen2020defining}.

\subsubsection{Compatibility Vulnerabilities and Weaknesses}
This category is related to vulnerabilities that prevent Ethereum from running normally on the developer machine. We find that the main root causes of this category are: (1) developers are not using the latest binaries/releases and (2) the hardware that is in use does not meet the minimum requirements.
As this category played only a minor role in the analyzed posts, we did not label the posts in detail, but decided to keep this for future work.

\subsubsection{Deployment and Configurations}

This category of vulnerabilities and weaknesses is caused by wrong configurations and weaknesses in deployment. 

\textbf{Improper configuration --- DL\#1}
Wrong or improper configuration of the smart contract application tool-chain can result in weaknesses, errors, or vulnerabilities, which applies even if the contract itself is free of vulnerabilities. 

\textbf{Violating contract size limit --- DL\#2.}
In Ethereum, limits are imposed by the gas consumed for the transaction. While there is no exact size limit there is a block gas limit and the amount of gas provided has to be within that limit. When deploying a contract, there are three factors to consider: (1) an intrinsic gas cost, (2) the cost of the constructor execution, and (3) the cost of storing the bytecode. The intrinsic gas cost is static, but the other two are not. The more gas consumed in the constructor, the less is available for storage. Normally, the vast majority of gas consumed is based on the size of the contract. If the contract size is large, the consumed gas can get close to the block gas size limits, preventing the deployment of the contract.

\begin{table}[htbp]
\centering
\caption{Authentication \& authorization vulnerabilities and weaknesses Attribute}
\label{table:Authenticationattribute}
\begin{tabular}{||p{2 cm} |p{6 cm}||} 
 \hline
 Attribute & Values\\ [0.5ex] 
 \hline
 Authentication \& authorization vulnerabilities and weaknesses & 	 unauthorized function call, wrong permissions, lack of access control, signature issues. \\
\hline
\end{tabular}
\end{table}

\subsubsection{Authorization and Authentication}
The following vulnerabilities and weaknesses directly affect the security of a smart contract and could enable attacks/exploits.

 \textbf{Lack of access control management --- SV\#1.} Access control is an essential element to the security of a smart contract. Based on the privileges of each client/contractor party, there have to be strict rules implemented in the contract that enforce access control. 
 
 For example, the contract in Listing~\ref{lst:access} is trying to provide functionality to whitelist addresses. The original function in line 10 does not have any access restrictions, meaning any caller can whitelist addresses. The modified version in line 15 only allows the contract owner to do so.
\begin{lstlisting}[language=Python, label={lst:access}, caption=Lack of access control management]
contract WHITELIST {
    address owner; //set during the first call
    modifier isOwner() {
        require(msg.sender == owner);
        _;
    }
    // insecure
    function enableWhitelist(address address) {
        //Whitelist an address
    }
    // secure
    function enableWhitelist(address address) external isOwner {
        //Whitelist an address
    }
}
\end{lstlisting}

\textbf{Authorization via tx.origin --- SV\#2}
tx.origin is a global variable in Solidity which returns the address of the account that sent the transaction. Rather than returning the immediate caller, tx.origin returns the address of the original sender (i.e., the first sender who initiated the transaction chain). It can make the contract vulnerable, if an authorized account calls into a malicious contract. Therefore, a call could be made to the vulnerable contract that passes the authorization check as tx.origin returns the original sender of the transaction, which in this case is the authorized account.

\textbf{Signature based vulnerabilities --- SV\#3}
These vulnerabilities are introduced as a result of insufficient signature information or weaknesses in signature generation and verification. Those include but not limited to: 
\begin{itemize}
    \item Lack of proper signature verification: One example can be relying on \emph{msg.sender }for authentication and assuming that if a message is signed by the sender address, then it has also been generated by the sender address. Particularly in scenarios where proxies can be used to relay transactions, this can lead to vulnerabilities. For more information, we refer to SWC-122\footnote{https://swcregistry.io/docs/SWC-122}
    \item Missing Protection against Signature Replay Attacks: To protect against Signature Replay Attacks, a secure implementation should keep track of all processed message hashes and only allows new message hashes to be processed. Without such control, it would be possible for a malicious actor to attack a contract and receive message hashes that were sent by another user multiple times. For more information, we refer to SWC-121\footnote{https://swcregistry.io/docs/SWC-121}
    \item Signature Malleability: The implementation of a cryptographic signature system in Ethereum contracts often assumes that the signature is unique, but signatures can be altered without the possession of the private key and still be valid. Valid signatures could be created by a malicious user to replay previously signed messages. For more information, we refer to SWC-117\footnote{https://swcregistry.io/docs/SWC-117}
\end{itemize}

\subsubsection{Dependency and Upgradability Vulnerabilities and Weaknesses}
Dependencies of and upgrades to smart contracts can lead to a number of issues in smart contracts. We describe one of these weaknesses/vulnerabilities below.

\textbf{Insecure contract upgrading --- UV\#1.}
There are two ways to upgrade a contract. First, to use a registry contract that keeps track of the updated contracts, and second, to split the contract into a logic contract and a proxy contract so that the logic contract is upgradable while the proxy contract is the same. Both approaches allow untrusted developers to introduce dependency vulnerabilities in the updated contract's logic, allowing attackers to modify the logic of the upgraded contract using its dependencies, e.g. another contract. This vulnerability was also reported in \cite{atzei2017survey} and \cite{chen2020survey}.

\subsubsection{Interface Vulnerabilities and Weaknesses}
This category describes weaknesses resulting in incorrect display of results.
However, in these cases the smart contracts actually worked and produced the expected outcomes.
The weaknesses were instead found in other applications that displayed the results.
As these applications are not part of the smart contract, and there was no apparent code weakness in the contracts, we did not further investigate these kind of weaknesses. 
\subsection{Are the frequency distributions of vulnerabilities similar across all studied data sources? (RQ2)}
\label{sec:resfreqt}
To answer RQ2, we analyzed the frequency distributions of the defined eleven vulnerability categories across the four data sources (i.e., Stack Overflow GitHub, CVE, and SWC). The frequency distribution of the defined categories is shown in Figure~\ref{fig:frequency} for each data source. 

\begin{figure*}[htbp]
\centering
    \subfloat[][Stack Overflow]{\resizebox{0.25\textwidth}{!}{
        \begin{tikzpicture}
    \begin{axis}[
        ybar,
        ymin=0,
        width  = 10cm,
        height = 8cm,
        bar width=14pt,
        ylabel={Percentage},
        xticklabel style={rotate=90},
        xtick = data,
        table/header=false,
        table/row sep=\\,
        xticklabels from table={\footnotesize
          CoV\\\footnotesize CV\\\footnotesize DL\\\footnotesize DV\\
          \footnotesize
          IB\\\footnotesize LV\\\footnotesize SV\\\footnotesize SV-DF\\
          \footnotesize
          SV-SC\\\footnotesize TV\\\footnotesize UV\\          }{[index]0},
        enlarge y limits={value=0.2,upper}
    ]
    \addplot table[x expr=\coordindex,y index=0]{1.83
\\29.02\\10.46\\16.99\\12.42\\14.25\\3.53\\5.1\\1.05\\1.83\\3.53\\};  \pgfplotsinvokeforeach{0,3,4,7,8,11}{\coordinate(l#1)at(axis cs:#1,0);}
    \end{axis}
    
\end{tikzpicture}
    }}
    \subfloat[][GitHub]{\resizebox{0.25\textwidth}{!}{
         \begin{tikzpicture}
    \begin{axis}[
        ybar,
        ymin=0,
        width  = 10cm,
        height = 8cm,
        bar width=14pt,
        ylabel={Percentage},
        xticklabel style={rotate=90},
        xtick = data,
        table/header=false,
        table/row sep=\\,
        xticklabels from table={\footnotesize
          CoV\\\footnotesize CV\\\footnotesize DL\\\footnotesize DV\\
          \footnotesize
          IB\\\footnotesize LV\\\footnotesize SV\\\footnotesize SV-DF\\
          \footnotesize
          SV-SC\\\footnotesize TV\\\footnotesize UV\\
          }{[index]0},
        enlarge y limits={value=0.2,upper}
    ]
    \addplot table[x expr=\coordindex,y index=0]{4.36\\20.03\\ 2.18\\17.65\\10.91\\10.11\\1.388\\16.07\\7.53\\5.35\\4.36\\};
    \pgfplotsinvokeforeach{0,3,4,7,8,11}{\coordinate(l#1)at(axis cs:#1,0);}
    \end{axis}
    
\end{tikzpicture}
    }}
    \subfloat[][CVE]{\resizebox{0.25\textwidth}{!}{
         \begin{tikzpicture}
    \begin{axis}[
        ybar,
        ymin=0,
        width  = 10cm,
        height = 8cm,
        bar width=14pt,
        ylabel={Percentage},
        xticklabel style={rotate=90},
        xtick = data,
        table/header=false,
        table/row sep=\\,
        xticklabels from table={\footnotesize
          CoV\\\footnotesize CV\\\footnotesize DL\\\footnotesize DV\\
          \footnotesize
          IB\\\footnotesize LV\\\footnotesize SV\\\footnotesize SV-DF\\
          \footnotesize
          SV-SC\\\footnotesize TV\\\footnotesize UV\\
          }{[index]0},
        enlarge y limits={value=0.2,upper}
    ]
    \addplot table[x expr=\coordindex,y index=0]{0\\0.19\\0\\2.88\\0\\5.96\\1.34\\89.61\\0\\0\\0\\};
  
    \pgfplotsinvokeforeach{0,3,4,7,8,11}{\coordinate(l#1)at(axis cs:#1,0);}
    \end{axis}
\end{tikzpicture}
    }}
    \subfloat[][SWC]{\resizebox{0.25\textwidth}{!}{
         \begin{tikzpicture}
    \begin{axis}[
        ybar,
        ymin=0,
        width  = 10cm,
        height = 8cm,
        bar width=14pt,
        ylabel={Percentage},
        xticklabel style={rotate=90},
        xtick = data,
        table/header=false,
        table/row sep=\\,
        xticklabels from table={\footnotesize
          CoV\\\footnotesize CV\\\footnotesize DL\\\footnotesize DV\\
          \footnotesize
          IB\\\footnotesize LV\\\footnotesize SV\\\footnotesize SV-DF\\
          \footnotesize
          SV-SC\\\footnotesize TV\\\footnotesize UV\\
          }{[index]0},
        enlarge y limits={value=0.2,upper}
    ]
    \addplot table[x expr=\coordindex,y index=0]{21.05\\15.78\\7.89\\10.52\\5.26\\5.26\\0\\0\\31.57\\0\\2.63\\};
    \pgfplotsinvokeforeach{0,3,4,7,8,11}{\coordinate(l#1)at(axis cs:#1,0);}
    \end{axis}

\end{tikzpicture}
    }}
    \caption{Distribution of SC vulnerability categories for each of the studied data sources.}
    \label{fig:frequency}
    \end{figure*}
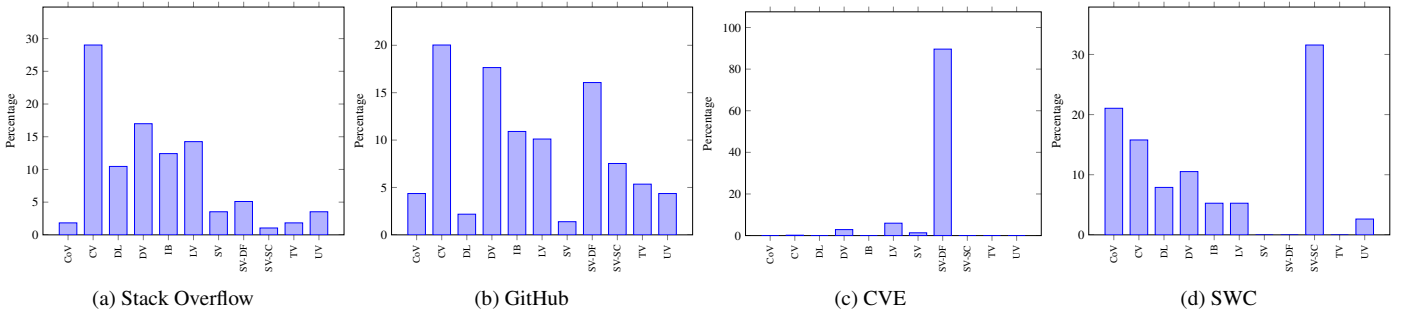 
\begin{tcolorbox} The resulting frequency distribution shows that the Language specific coding category and the Structural data flow category are the most common vulnerability categories in Ethereum smart contracts \end{tcolorbox}

The language specific coding category is dominant on both Stack Overflow and GitHub. In contrast, structural data flow vulnerabilities are most frequent on CVE and SWC. Additionally, almost 80\% of the reported structural data flow vulnerabilities in the NVD database (CVEs) are integer overflow/ underflow vulnerabilities. 
Interestingly, issues on StackOverflow and GitHub appear to have similar frequency distributions.
However, these are not statistically significant.
 \subsection{What impact do the different categories of smart contract vulnerabilities and weaknesses have? (RQ3)}
\label{sec:resImpact}
In this section, we present the main impacts of vulnerabilities and bugs in smart contracts, thus answering RQ3. We unify all the proposed impact categories in the literature (i.e., \citep{chen2020defining,zhang2020framework}), and  present a thorough classification scheme of vulnerability impacts on Ethereum smart contracts. We followed a similar approach to what we did in RQ1.  
The definitions of the final impact categories are depicted in Table~\ref{table:impacts} and Table~\ref{table:impacttable2}.
Furthermore, Table~\ref{table:mapping_id} shows how the impact categories in literature related to our classification.
Note that IP5 (in the second-last row) does not relate to any of our categories, as the category describes smart contracts that function as intended, something that we did not include in our classification.

\begin{tcolorbox}Our mapping shows that the impact of smart contract vulnerabilities and code weaknesses on certain aspects has not been examined in detail. For instance, the impact on the development process and the productivity of a software development team. Additional research in this area can quantify the extent to which the vulnerabilities impact smart contract development, the developing team, and the development of decentralized applications based on smart contracts.\end{tcolorbox}

\begin{table}[htbp]
\centering
\caption{Classification Scheme of Impacts (I-D1)}
\label{table:impacts}
\begin{tabular}{||p{2.2 cm} | p{4.8cm}| p{.4 cm} ||} 
 \hline
 Impact & Description & App.\\ 
 \hline\hline
Unexpected behavior & Contract behaves abnormally, e.g., generating incorrect output. & UB\\

Unwanted functionality & Contract executes wrong functionality because of wrong logic. & UF\\

Long response time & Long runtime of a smart contract to any input without providing the desired output.& LRT\\

Data Corruption & Data becomes unreadable, unusable or inaccessible, and unexpected output is generated.& DC\\

Memory disclosure & Problems in the memory storage of the smart contract. & MF \\

Poor performance & Non-optimal resource usage in terms of gas and time. & PPC\\

Unexpected stop & Unexpected exit and execution stop at the point of triggering the vulnerability in the code.& USP
\\ 
\hline
\end{tabular}
\end{table}

\begin{table}[htbp]
\centering
\caption{Classification Scheme of Impacts (I-D2)}
\label{table:impacttable2}
\begin{tabular}{||p{2 cm} | p{5 cm} | p{.5 cm}||} 
 \hline
 Impact & Description & App.\\ 
 \hline\hline
Information disclosure & When a code weakness or vulnerability is exploited, sensitive information is exposed to an actor not explicitly authorized to see it.& ID\\

Lost Ether or assets& An exploited code weakness can lead to unauthorized actors taking over the Ether of the contract and losing it.& LEA\\

Locked Ether& In the situation of triggering code weaknesses in a contract, one can lose access to the contract and lock the Ether in it without having access to it again. & LE\\

Lost control over the contract & By exploiting code weaknesses or vulnerabilities, an unauthorized actor can take over the contract. & LC\\
\hline
\end{tabular}
\end{table}

\begin{table*}[htbp]
\centering
\caption{Mapping literature-based Impact classifications to I-D. \emph{D1}: refers to the Impact on software product. Note: IP2 belongs to D1 and D2. {$\subset$}* indicates the corresponding category in our own classification is a subset of the corresponding category in the literature marked by *. *{$\subset$} means the category in the literature is a subset of our proposed category.}
\label{table:mapping_id}
\begin{tabular}{|l||*{11}{c|}}\hline
\backslashbox{Literature}{Ours}

&\makebox[1.5em]{UB}&\makebox[1.5em]{UF}&\makebox[1em]{LRT}
&\makebox[1em]{DC}&\makebox[1.2em]{MF}&\makebox[2em]{PPC}&\makebox[1.2em]{USP}&\makebox[1.2em]{ID}&\makebox[1.2em]{LEA}&\makebox[1.2em]{LE}&\makebox[1.2em]{LC}\\\hline\hline

 \tikzmark[xshift=-8pt,yshift=1ex]{x}Functionality*. \citep{zhang2020framework}&   & = &&  & & & & && &\tikzmark[xshift=3.5em]{a} \\ \hline 

Performance* \citep{zhang2020framework}&   &  &&  & &= & & && &\\\hline
Security \citep{zhang2020framework}& \textbf{$\subset$}*  & \textbf{$\subset$}* &\textbf{$\subset$}*& \textbf{$\subset$}* &\textbf{$\subset$}* &\textbf{$\subset$}* &\textbf{$\subset$}* &\textbf{$\subset$}* &\textbf{$\subset$}*&\textbf{$\subset$}* &\textbf{$\subset$}*\\\hline
Serviceability *\citep{chen2020defining}& \textbf{$\subset$}*  & \textbf{$\subset$}* &\textbf{$\subset$}*& \textbf{$\subset$}* &\textbf{$\subset$}* &\textbf{$\subset$}* &\textbf{$\subset$}* & && &\\\hline
IP1  unwanted behaviors* \citep{chen2020defining} & =  &  &&  & & & & && &\\\hline
IP3 non-exploitable UB* \citep{chen2020defining} & *\textbf{$\subset$}  &  &&  & & & & && &\\\hline
IP4 Errors outside program call \citep{chen2020defining} &   &* \textbf{$\subset$}  &&  & & & & && &\\\hline
 \tikzmark[xshift=-8pt,yshift=1ex]{y}IP5 No errors* \citep{chen2020defining} &   &  &&  & & & & && &\tikzmark[xshift=3.5em]{a} \\\hline
  IP2 UB without losses* \citep{chen2020defining} & *\textbf{$\subset$} &  &&  & & & & & &*\textbf{$\subset$}& \\\hline

\end{tabular}
\drawbrace[brace mirrored, thick]{x}{y}
\annote[left]{brace-4}{I-D1}
\end{table*}

The analysis of the impacts of vulnerabilities and code weaknesses in smart contracts shows that unexpected stop is the most prevalent impact category among all proposed categories. It is caused primarily by vulnerabilities in the Language Specific Coding category.
The second most prevalent impact category is unexpected functionality. This mostly happens in smart contracts when the transferred gas amount does not match the expected amount expected from the logic of the code, or when incorrect amounts of Ether are transferred.

\section{Discussion}
\label{sec:discussion}

In the following, we discuss our findings in terms of implications and relation to existing work.

We find a substantial number of vulnerabilities and weaknesses being discussed in social coding platforms and existing vulnerability repositories.
This clearly shows that this is an important topic to study and analyze.
Because of the unique characteristics of smart contracts, e.g., immutability and gas consumption, it is important to make sure that vulnerabilities and code weaknesses with severe impacts are fixed or even detected before deploying the contract to the blockchain.

As demonstrated in our findings, existing classifications either focus on a single dimension of smart contract vulnerabilities, such as the error source, or mix multiple dimensions in a single classification.
Our mapping unifies these different dimensions to some extent and shows how different classification schemes relate.

In addition to mixing dimensions, the majority of existing classification schemes for smart contract vulnerabilities include broad categories, such as security and availability, to which many vulnerabilities can be assigned. As such, these categories do not support reasoning about the included vulnerabilities, which is an important quality criterion for classifications \cite{ralph18}.
Furthermore, broad categories might prevent orthogonality of the categories, i.e., that a single vulnerability fits into a single category only.
In the example mentioned above, i.e., security and availability, many vulnerabilities can lead to negative effects on both, and thus could be labeled both.
Due to the use of attribute-value pairs \cite{seacord2005structured}, we believe that our unified classification avoids this issue.

The frequency distributions discussed in Section~\ref{sec:resfreqt} show notable differences between the frequencies of found vulnerabilities across different data sources. This, once again, highlights that focusing on a single source biases the resulting study.
It further demonstrates that established databases such as CVE and SWC do not reflect well the topics discussed in public coding platforms such as Stack Overflow or GitHub.
On the latter platforms, we observe specifically that developers seem to have a poor understanding of pre-defined functions such as \emph{view} and \emph{pure}, and that it is hard for them to cope with continuous changes and updates in Solidity and its documentation.
On the one hand, this finding suggests that tools for verification and analysis of smart contracts is of high importance, especially focusing on the prevalent vulnerability categories in our classification.
On the other hand, the observed frequencies might only be a symptom of the technology maturity. Hence, these issues might become less prevalent once Solidity matures and updates become less frequent.

Based on our findings, we can provide a number of recommendations to researchers and practitioners in order to improve smart contract development.

First, available static detection tools must urgently target the defined categories of vulnerabilities and code weaknesses. Our breakdown of the frequencies at which the different categories occur can help prioritizing this development to target the most important categories first.
Second, there is a need to define coding best practices of smart contracts and make them available to developers. Coding guidelines are available for many programming languages and technologies, and can help improving quality and reducing code weaknesses and unexpected gas consumption. Based on the analyzed data from social coding platforms, we believe that this list can include, e.g., avoiding multidimensional arrays, or arrays in general, if possible; carefully checking gas consumption amounts; recomming specific safe libraries such as \emph{SafeMath}~\cite{Safemath} to avoid common pitfalls such as underflows; and avoiding substantial sub-contract creation.
Finally, we see a need for action from the Solidity team, in particular when it comes to clearly documenting existing libraries and code weaknesses they can resolve, as well as clearly defining gas requirements for patterns, functions, declaration types, creating subcontracts, upgrading contracts, and other language artifacts and operations.
Such documentation could substantially contribute to a reduction in smart contract vulnerabilities and weaknesses in Solidity/Ethereum.

\section{Threats to Validity}
\label{sec:threats}
In the following, we discuss threats to the validity of our findings according to internal and external validity, as well as reliability.

\subsection{Internal Validity}
Card sorting was used as the main method to categorize and label the posts in this work.
This method is subjective and open to bias, hence two experts verified the manual labeling twice independently in order to mitigate this threat.
In case of any disagreement, the two experts would discuss it until reaching to a consensus.

An additional threat to internal validity lies in the keyword search employed during data collection.
For instance, to collect Q\&A posts from Stack-Overflow, we used the tags ``Smart contract'', ``Ethereum'', and ``Solidity'', while for GitHub we used ``smart contract'', ``Solidity'', and ``Ethereum''.
Also, a part of the collected data with insufficient information or with information that is not related to smart contract vulnerabilities and code weaknesses written in Solidity were excluded from the analysis.
This search and filtering process might have led to exclude relevant data that could have led to further insights.

As trustworthiness of the collected posts can be a source of noise in the data, we removed negative voted questions from the dataset.
This might have resulted in a systematic under-representation of certain types of vulnerabilities/bugs.
Given the maturity of Solidity, we deemed this step warranted to allow for data of sufficiently high quality.

\subsection{External Validity}
A threat to external validity is the continuous updates of Ethereum using the hard fork.
More improvements are being added to Ethereum with each hard fork such as the EIPs \footnote{https://github.com/ethereum/EIP} that ensures the energy efficiency of the proof-of-work.
Furthermore, many new features are added to Solidity newer versions.
Therefore, new contract vulnerabilities and bugs may be introduced and some others may be resolved.
This means that our classification might not generalize to newer (or older) versions of Solidity.

Focusing on Solidity and the Ethereum blockchain only limits the external validity of the results, as there are other blockchains and other smart contract languages that could have yielded further or different vulnerabilities/bugs.
We focus on Ethereum since it is the second-largest blockchain, and the largest blockchain that supports smart contracts (written in Solidity).
However, additional studies into other technologies could be a valuable addition to our findings, and our results might not generalize to other blockchains or smart contract languages.

\subsection{Reliability}
To ensure reliability of the results, we described in detail the data collection and analysis process.
We systematically calculated inter-rater reliability coefficients in iterative rounds, to ensure sufficiently clear categories and agreement among two expert raters.
Finally, we published the full dataset~\cite{dataset}.
Overall, these steps should ensure reliability and enable replication of our study. 
\section{Conclusion}
\label{sec:conclusion}
Due to the immaturity of blockchain technology, vulnerabilities in smart contracts can have severe consequences and result in substantial financial losses. 
Hence, it is essential to understand vulnerabilities and code weaknesses of smart contracts.
However, there exist several shortcomings in existing classifications, i.e., they focus on single dimensions, mix dimensions, propose too broad categories, or rely on single data source omitting important sources for vulnerabilities.
To address this gap, we extracted smart contract vulnerabilities written in Solidity from a number of important data sources, classified them in terms of error source and impact, and related them to existing classifications in literature.
Our findings show that language-specific coding and structural data flow are the dominant categories of vulnerabilities in Solidity smart contracts.
We also find that many vulnerabilities and code weaknesses are similar to known issues in general purpose programming languages, such as integer overflow and erroneous memory management. 
However, the immaturity and rapid evolution of the technology and the Solidity language, and the added concept of gas furthermore adds vulnerabilities and code weaknesses, and increases the risk of attacks and financial losses.
Interestingly, we find that the frequency at which the different categories occur differs widely across data sources, indicating that they should not be viewed in isolation.
 
Our classification scheme is a further step to standardize and unify vulnerability analysis in smart contracts.
This can support researchers in building tools and methods to avoid, detect, and fix smart contract vulnerabilities in the future.

Specifically, we see a number of directions for future research.
First, future studies should investigate whether our classification scheme can be generalized by investigating other smart contracts in other blockchain networks, e.g., Hyperledger, Stellar, and Openchain. Potentially, our classification needs to be modified to fit into other networks or languages. Similarly, the classification might have to be adapted as Solidity evolves and as new languages are developed for Ethereum smart contracts.
Second, more dimensions and characteristics of the studied vulnerabilities and code weaknesses can be explored in the future. For instance, patterns among the extracted vulnerabilities and code weaknesses could be abstracted, as well as the evolution over time. For each of the categories, code metrics and detection tools can be explored.
Finally, it should be investigated in which ways the defined vulnerabilities and code weaknesses can be exploited, as well as what the impact of these exploits will be. This can be accomplished by developing automated scripts or manually devising such exploits. 
\section*{Acknowledgment}
The authors would like to thank {Mohammad Alsarhan}, a security expert, for participating in the card sorting and inter-rater agreement discussions.
This work was supported by the {Icelandic} Research Fund (Rannís) grant number 207156-051.

\bibliographystyle{abbrvnat}

\end{document}